\documentclass[12pt,letterpaper]{article}

\usepackage{amsmath,amssymb,calc}
\usepackage{graphicx}

\newcommand\be{\begin{equation}}
\newcommand\ee{\end{equation}}
\newcommand{\bea}{\begin{eqnarray}}
\newcommand{\eea}{\end{eqnarray}}

\newcommand{\la}{\langle}
\newcommand{\ra}{\rangle}

\newcommand{\nn}{\nonumber}
\newcommand{\pd}{\partial}

\def\id{\protect{{1 \kern-.28em {\rm l}}}}
\def\idp{\protect{{\rm 1 \kern-.28em {\rm l}}}}

\def\unit{\relax{\rm 1\kern-.26em I}}

\def\id{\protect{{1 \kern-.28em {\rm l}}}}

\begin{document}

\begin{titlepage}
\mbox{}
\vspace{-0.5cm}
\begin{center}
\hfill QMUL-PH-07-05 \\
\hfill Imperial/TP/07/RR/01 \\
\vskip 6mm

{\Large {\bf Metastable SUSY Breaking and Supergravity at Finite Temperature\\[3mm] }}

\vskip 3mm

\vskip .1in

{\bf Lilia Anguelova$^{a}$\footnote{{\tt l.anguelova@qmul.ac.uk}},
Riccardo Ricci$^{b,c}$\footnote{{\tt r.ricci@imperial.ac.uk}} and Steven
Thomas$^{a}$\footnote{{\tt s.thomas@qmul.ac.uk}}}

\vskip 2mm
$^a${\small {\em Center for Research in String Theory}\\
{\em Department of Physics, Queen Mary, University of London}\\
{\em Mile End Road, London, E1 4NS, UK}}\\[2mm]
$^b${\small {\em Theoretical Physics Group, Blackett Laboratory}\\
{\em Imperial College, London, SW7 2AZ, UK}}\\[2mm]
$^c${\small {\em The Institute for Mathematical Sciences,}\\
{\em Imperial College, London, SW7 2PG, UK}}\\

\vskip 3mm

\end{center}

\vskip .3in

\begin{center} {\bf ABSTRACT }\end{center}

\vspace{-0.8cm}
\begin{quotation}\noindent

We study how coupling to supergravity affects the phase structure of a
system exhibiting dynamical supersymmetry breaking in a metastable vacuum. More
precisely, we consider the Seiberg dual of SQCD coupled to supergravity at
finite temperature. We show that the gravitational interactions decrease the
critical temperature for the second order phase transition in the quark
direction, that is also present in the global case. Furthermore, we find that,
due to supergravity, a new second order phase transition occurs in the meson 
direction, whenever there is a nonvanishing constant term in the superpotential. 
Notably, this phase transition is a necessary condition for
the fields to roll, as the system cools down, towards the metastable susy
breaking vacuum, because of the supergravity-induced shift of the
metastable minimum away from zero meson vevs. 
Finally, we comment on the
phase structure of the KKLT model with uplifting sector given by the Seiberg
dual of SQCD.

\end{quotation}
\vfill

\end{titlepage}

\eject

\tableofcontents

\section{Introduction}
Understanding how supersymmetry breaking occurs is a major problem
on the road to connecting the underlying supersymmetric theory
with the observed world\footnote{For a recent review of
supersymmetry breaking see \cite{Intriligator:2007cp}.}. The idea,
that supersymmetry can be broken due to dynamical effects
\cite{Witten}, has long been considered phenomenologically very
promising, since it naturally leads to a large hierarchy between
the Planck and the susy breaking scales. However, dynamical
supersymmetry breaking has turned out to be quite difficult to
implement in a supersymmetric gauge theory. The reason is that
only rather complicated examples \cite{exDSB} satisfy the strict
conditions, necessary for the absence of a global supersymmetric
vacuum.

It was realized recently \cite{ISS} that the situation changes
dramatically, if one abandons the prejudice that the
phenomenologically relevant vacuum has to be a global, and not
just local, minimum of the effective potential. In this case, one
can relax the requirement that the theory lacks a global
supersymmetric vacuum and search for models with meta-stable,
sufficiently long-lived, susy breaking vacua. From this new point
of view, models with non-zero Witten index and without a conserved
$U(1)$ R-symmetry can be considered phenomenologically viable
for supersymmetry breaking. The spectrum of susy breaking
theories is then significantly enriched. In particular, as shown
in \cite{ISS}, meta-stable dynamical supersymmetry breaking occurs
even in ${\cal N}=1$ SQCD with $SU(N_c)$ gauge group and $N_f$
massive fundamental flavors. This can be established by going to
the Seiberg (magnetic) dual description of this theory, where
supersymmetry is broken at tree-level.\footnote{We will review
more details on that in the next Section.} For convenience, we
will call this the ISS model from now on. During the last year,
many more examples of meta-stable dynamical susy breaking were
found in various phenomenologically appealing settings
\cite{msDSB}. Progress was also made on understanding the
embedding of those field-theoretic models in string/M-theory
\cite{msDSBstr}.

However, once we consider phenomenology in a local, instead of a
global, minimum of the zero-temperature effective potential, the
following question arises. How natural is it for the
high-temperature system, that is the early Universe, to end up in
the metastable state after cooling down? To address this question,
the recent works \cite{ACJK, FKKMT} studied the ISS model at
finite temperature. They found that the metastable vacuum is
thermodynamically preferable compared to the supersymmetric global
ones.\footnote{There are $N_c$ of them as we review in Section
\ref{ISSrev}.} Although their conclusions agree, their approaches
are different and, in a sense, complimentary. In \cite{ACJK}, they
consider a path in field space, which extrapolates between the susy
breaking vacuum and a global vacuum, and construct the effective
potential along this path. Using this they show that, even if at high
temperature the system starts at a susy vacuum, it will end up in the
metastable one as it cools down. On the other hand, \cite{FKKMT} studies
in great detail the phase structure around the origin of field space,
which is a local minimum of the nonzero temperature potential. They
assume that at high temperature the quark and meson fields of the ISS
model are localized near this point, which is reasonable
since the number of light degrees of freedom at the
origin is largest and hence this state maximizes the entropy. With
this starting point, \cite{FKKMT} investigates the phase structure
of the free-energy as the temperature decreases and finds out that
there is a critical temperature, $T_c^Q$, for a second order phase
transition in the quark direction (that is, the direction towards
the metastable minimum). On the other hand, in the meson direction
(i.e., the direction towards a global susy vacuum) they find that
only a first order phase transition can occur, at temperature
smaller than $T_c^Q$, and that it is quite suppressed by a high
potential barrier.\footnote{ Reference \cite{CFW} discusses
in detail the suppression of this transition in a class of
O'Raireataigh models.}

We follow the approach of \cite{FKKMT} for the ISS model coupled to 
supergravity. This is certainly necessary for more realistic cosmological 
applications of the idea of dynamical supersymmetry breaking in a metastable 
vacuum. We will compute the one-loop effective potential at finite temperature 
by using the results of \cite{BG1, BG2} for chiral multiplets coupled to 
supergravity. The nonrenormalizability of the latter theory is not an issue as 
it is supposed to be viewed as an effective low-energy description only, not 
as a fundamental theory. For more details on one-loop (albeit at $T=0$) calculations 
in supergravity coupled with various matter multiplets, see \cite{OneLoop}. 
However, we should note that the considerations of \cite{BG1, BG2} treat the 
zero-temperature classical supergravity contribution $V_0$ to the $T\neq 0$ effective 
potential as an effective potential itself. In other words, the $T=0$ loop 
corrections are viewed as already taken into account in the standard sugra potential 
$V_0$. So it might seem conceivable that these results may be affected (despite susy 
being broken) by regularization subtleties similar to those at $T=0$. In the latter 
case, a regularization compatible with supersymmetry was developed in the last five 
references of \cite{OneLoop} and was shown rather recently in \cite{GN} to sometimes have 
an impact on quantities of phenomenological interest, like the flavor-changing 
neutral currents. For the trivial K\"{a}hler potential, that we will need, this is not 
the case. Nevertheless, it is important, although going well beyond the scope of our 
paper, to address this issue in full generality at $T\neq 0$. Another remark is due.
Every gravitational system exhibits 
instability under long-wave length gravitational perturbations \cite{Jeans}; this Jeans 
instability occurs also at finite temperature \cite{GPY}. While it is certainly of great 
importance for structure formation in the early Universe, it is a subleading effect 
on cosmological scales on which the Universe is well-approximated by a homogeneous fluid. 
So we will limit ourselves to considering the leading effect, by using the formulae 
of \cite{BG1, BG2} for the effective potential, and will not address here the Jeans 
instability.

We will show that the supergravity corrections decrease the critical temperature 
for a second order phase transition in the quark direction, $T_c^Q$, for
any $N_c$ and $N_f$. While this is only a small quantitative
difference with the rigid case, in the meson branch a significant
{\it qualitative} difference can occur. The reason is that in the relevant 
field-space region there are no contributions to the tree-level
meson masses in the rigid limit and so the supergravity
corrections are the leading ones. As a result, it turns out that, whenever 
the superpotential contains a nonvanishing constant piece $W_0$, 
there is a second order phase transition in the meson direction at
a temperature $T_c^{\varphi}$, smaller than $T_c^Q$. However, this 
is not a phase transition towards any of the global supersymmetric 
minima, as it occurs at nonvanishing quark vevs. Notably though, our 
new phase transition is precisely what is needed for the system to 
roll towards the metastable vacuum since, due to the supergravity 
interactions, the latter is shifted away from the origin of the meson 
direction whenever $W_0\neq 0$, as shown in \cite{AHKO}.

Considering the ISS model plus supergravity is, in fact,
the first step towards a full investigation of the phase structure of the KKLT
scenario \cite{KKLT} with ISS uplifting sector at finite temperature. It was
already argued in \cite{DPP, AHKO} that metastable susy breaking provides a
natural way of lifting the AdS KKLT vacuum to a de Sitter one, avoiding the
problems encountered previously in the literature. Recall that the original
proposal was to introduce anti-D3 branes, which break supersymmetry explicitly,
whereas the later idea to use nonvanishing D-terms \cite{BKQ} turned out to be
quite hard to realize \cite{CFNO}. Studying the phase structure of the KKLT-ISS
system is a big part of our motivation. However, in this case the computations
become much more technically challenging. We make the initial step by showing that
the origin of the ISS field space is no longer a local minimum of the
temperature-dependent part of the effective potential. The shift of the high
temperature minimum away from the ISS origin is related to the vev of the KKLT
volume modulus.

This paper is organized as follows. In Section \ref{ISSrev} we review
necessary background material about the ISS model. In Section \ref{VT} we compute
the one-loop temperature-dependent contribution to the effective potential or,
equivalently, the free energy of the ISS model coupled to supergravity. To
achieve that, in Subsection 3.1 we derive the mass matrices for both quark and
meson vevs nonzero, that are coming from the F-terms; in Subsection 3.2 we
take into account the D-terms. In Section 4 we expand the general results of
Section 3 in terms of the small parameter $M_P^{-1}$, where $M_P$ is the
Planck mass. This allows us to read off the leading supergravity corrections
to the rigid theory, considered in \cite{FKKMT}. In Section 5 we compute the
critical temperature for a second order phase transition in the quark direction
to ${\cal O}(M_P^{-2})$. In Section 6 we show that there is also a second order
phase transition in the meson direction and estimate the critical temperature for
it. In Section 7 we consider the KKLT model with ISS uplifting sector and argue
that the origin of the ISS field space is no longer a local minimum of the high
temperature effective potential. The shift of the minimum away from this origin
is determined by the vev of the KKLT volume modulus. In Section 8 we discuss the
implications of our results for the phase structure of the ISS model coupled
to supergravity and for the end point of this system's evolution at low temperature. 
We also outline open problems. Finally, in Appendix A we give some useful formulae 
for mass matrices near the origin of field space and in Appendix B we show that 
no new supersymmetric minima appear in a small neighborhood of the origin in the 
ISS model coupled to supergravity, in the field directions of interest.

\section{ISS model} \label{ISSrev}

It was argued in \cite{SeibDual} that SQCD with $SU(N_c)$ gauge group
and $SU(N_f)_L\times SU(N_f)_R$ flavor symmetry has a dual (magnetic)
description in terms of an $SU(N_f-N_c)$ gauge theory coupled to certain
matter fields. When the condition $N_f < 3N_c/2$ is satisfied
the magnetic theory is IR free. The matter content of the dual theory
comprises two chiral superfields $q$ and $\tilde{q}$, that are transforming
in the $({\bf \overline N_f},{\bf 1})$ and $({\bf 1},{\bf N_f})$ representations
of the flavor symmetry group respectively and are in the fundamental and
antifundamental representations of $SU(N_f-N_c)$ respectively (and so are
called quarks), and a
gauge-singlet chiral meson superfield $\Phi$ in the $({\bf N_f},{\bf
\overline N_f})$ flavor group representation. Hence the index structure of
the magnetic quark and meson fields is the following: $q_i^a$,
$\tilde{q}^{\bar{j}}_a$ and $\Phi^i_{\bar{j}}$ with $i=1,...,N_f$ and
$a=1,...,N\,,$ where $N=N_f - N_c$. For easier comparison with the literature,
we will also use the notation $N_m=N$ and $N_c=N_e$, implying in particular
that $N_f = N_m + N_e$. In terms of this notation the above condition for IR
free dual theory is $N_e > 2 N_m$. In the following we will only consider
this case.

The magnetic theory has the following tree-level superpotential:
\be \label{Wtree}
W = h \,{\rm Tr} \,q \Phi \,\tilde{q} - h \,\mu^2 \,{\rm Tr} \Phi \, .
\ee
The second term breaks the flavor group to its diagonal subgroup and
corresponds to a
quark mass term in the microscopic theory (i.e., the original $SU(N_c)$
gauge theory). The K\"{a}hler potential is the canonical one:
\be \label{KaP}
K = {\rm Tr} \,q^{\dagger} q + {\rm Tr} \,\tilde{q}^{\dagger} \tilde{q} +
{\rm Tr} \,\Phi^{\dagger} \Phi \, .
\ee

The magnetic description can be used to prove the existence of a
metastable vacuum, which breaks supersymmetry at tree-level \cite{ISS}.
Indeed, it is immediate to see that the F-term condition
\be
F_{\Phi_i^j}=h\left(q_i^a\tilde{q}^j_a-\mu^2\delta^j_i\right)=0
\ee
cannot be satisfied as the matrix $q_i^a\tilde{q}^j_a$ has at
most rank $N_m$ while $\delta^j_i$ has rank $N_f$. The moduli
space of metastable vacua can be parameterized as:
\be \label{metasm}
q=\left(%
\begin{array}{c}
 Q \\
  0 \\
\end{array}%
\right),\,\,\,\,\,
\tilde{q}^{\,T}=\left(%
\begin{array}{c}
  \tilde{Q} \\
 0 \\
\end{array}%
\right), \,\,\,\,\,
\Phi=\left(%
\begin{array}{cc}
  0 & 0 \\
  0 & \varphi \\
\end{array}%
\right) \, , \ee where $\varphi$ is an $(N_f - N)\times (N_f - N)$
matrix while $Q$ and $\tilde{Q}$ are $N\times N$ matrices
satisfying the condition $Q\tilde{Q}=\mu{\unit_{N\times N}}$. The
point of maximum global symmetry is at
\be \label{meta_sv}
\langle q_1 \rangle = \langle \tilde{q}_1^T \rangle = \mu \unit_N
\, , \qquad \langle q_2 \rangle = \langle \tilde{q}_2 \rangle = 0
\, , \qquad \langle \Phi \rangle = 0 \, ,
\ee
where we have denoted: $q^T\equiv (q_1, q_2)$ with $q_1$ and $q_2$
being $N\times N$ and $N\times (N_f-N)$ matrices respectively.
It will also be useful for the future to introduce the following
notation for the generic components of the $N_f \times N_f$ matrix
$\Phi$:
\be
\Phi=\left(%
\begin{array}{cc}
  \phi_{11} & \phi_{12} \\
  \phi_{21} & \phi_{22} \\
\end{array}%
\right) \, ,
\ee
where $\phi_{11}$ is $N_m\times N_m$, $\phi_{12}$ is $N_m \times N_e$,
$\phi_{21}$ is $N_e \times N_m$ and finally $\phi_{22}$ is an $N_e
\times N_e$ matrix.

The value of the scalar potential in each
metastable minimum in (\ref{metasm}) is:
\be
V_{\rm{min}}=(N_f-N) \,h^2\mu^4 \, .
\ee
Usually, when supersymmetry is spontaneously broken, the moduli
space of classical vacua is not protected against quantum corrections.
As a result, the quantum moduli space is typically smaller and one may
wonder whether any of the metastable vacua survive in it.
In this regard, it was shown in \cite{ISS} that the
classically flat directions around the maximally symmetric vacuum
(\ref{meta_sv}) acquire positive masses at one-loop through the
supersymmetric Coleman-Weinberg potential \cite{CW}. This metastable
minimum is therefore tachionic-free and from now on we will always mean
(\ref{meta_sv}), when we talk about a supersymmetry breaking vacuum.

In addition to the perturbative corrections that we just
discussed, there are also non-perturbative ones. Namely, gaugino
condensation in the magnetic gauge group $SU(N)$ induces the
Affleck-Dine-Seiberg superpotential \cite{ADS}:
\be \label{WADS}
W_{\rm{ADS}}=N\left(h^{N_f}{\det \Phi\over
\Lambda^{N_f-3N}}\right)^{1\over N} \, ,
\ee
where $\Lambda$ is
the UV cutoff of the magnetic theory, i.e. the scale above which
the magnetic description becomes strongly coupled and hence not
well-defined. Adding this dynamically generated contribution to
the classical ISS superpotential leads to $N_c$ {\it
supersymmetric} vacua, characterized by nonvanishing meson
vevs:\footnote{We denote by $\epsilon$ the quantity $\mu\,
\Lambda^{-1}$.}
\be \label{PhiADS}
h\langle\Phi^i_{\bar{j}}\rangle=\mu\,\epsilon^{{3N-N_f\over
N_c}}\,\delta^i_{\bar{j}}\,,\,\,\,\,\,\,\,\,\,\,\,\,\langle q
\rangle = \langle \tilde{q}^T \rangle = 0 \, ,
\ee
in agreement
with the Witten index \cite{Wittenind} of the microscopic theory.
The metastable vacuum can be made long-lived by taking $\epsilon$
parametrically small as in that case the tunnelling to the
supersymmetric vacuum is strongly suppressed.\footnote{Recall that
$N_f > 3 N$ and hence for very small $\epsilon$ the meson vev,
$\la \Phi \ra$, in (\ref{PhiADS}) becomes very large.}

Since the $ADS$ superpotential is suppressed by powers of the UV
cutoff, for small meson fields it is completely negligible
compared to the tree-level one, (\ref{Wtree}). So in the following
we will drop $W_{ADS}$ from our considerations, as we will study
the finite temperature effective potential only in a neighborhood
of the origin of field space.

\section{One-loop effective potential at nonzero $T$} \label{VT}
\setcounter{equation}{0}

In the present section we compute the one-loop effective potential
at finite temperature for the ISS model coupled to supergravity. Its
analysis in subsequent sections will enable us to deduce the phase
structure of this theory near the origin.

Let us start by recalling some generalities about the path-integral
derivation of the effective potential in a theory with
a set of fields $\{ \chi^I \}$. An essential step in that is to shift
$\chi^I$ by a constant background $\hat{\chi}^I$. Equivalently, we
expand the Lagrangian around a nonzero background, $\{ \hat{\chi}^I \}$,
for the fields. Using this expansion, one can derive with functional
methods a formula for the effective potential. The original derivation
of \cite{Jackiw} was only for zero-temperature renormalizable
field theory. The same kind of considerations apply also for finite
temperature and up to one-loop give \cite{DJ}:
\be
V_{eff} (\hat{\chi}) = V_{tree} (\hat{\chi}) + V_0^{(1)} (\hat{\chi})
+ V_T^{(1)} (\hat{\chi}) \, ,
\ee
where $V_{tree}$ is the classical potential, $V_0^{(1)}$ is the zero
temperature one-loop contribution, encoded in the Coleman-Weinberg
formula, and finally the temperature-dependent correction is:
\be
V_T^{(1)}(\hat{\chi}) = -\frac{\pi^2 T^4}{90}
\left( g_B+ \frac{7}{8} g_F \right)+ \frac{T^2}{24} \left[{\rm Tr}
M_s^2(\hat{\chi}) +3 {\rm Tr} M_v^2(\hat{\chi})+ {\rm Tr}
M_f^2(\hat{\chi}) \right] + {\cal O}(T)\label{Texpansion} \, .
\ee
For convenience, from now on we will denote this last expression simply
by $V_T$. Here $g_B$ and $g_F$ are the total numbers of bosonic and
fermionic degrees of freedom respectively\footnote{Note that this is
different from the number of fields. For example, for $N_B$ chiral
superfields the number of scalar degrees of freedom is $g_B = 2 N_B$.
In fact, below we will denote by $N_B$ the number of complex
scalars.}; ${\rm Tr} M_s^2$, ${\rm Tr}
M_v^2$ and ${\rm Tr} M_f^2$ are the coefficients of the quadratic terms
of scalar, vector and fermion\footnote{In (\ref{Texpansion}) the
quantity ${\rm Tr} M_f^2$ is computed summing over Weyl fermions.}
fields computed from the shifted classical potential or, in other words,
the mass matrices of those fields in the classical background
$\{\hat{\chi}^I\}$. The expansion (\ref{Texpansion}) is valid in the
high temperature regime, more precisely when all masses are much smaller
than the energy scale set by the temperature. The above result for the
one-loop effective potential at finite temperature was shown in
\cite{BG1} to also hold for coupling to supergravity.

We turn now to computing ${\rm Tr} M_s^2$, ${\rm Tr} M_v^2$
and ${\rm Tr} M_f^2$ for our case. The classical background $\hat{\chi}$
around which we will be expanding is:
\be
\la q_1 \ra = \la \tilde{q}_1 \ra = Q \,\unit_{N_m\times N_m} \, ,
\qquad \la \phi_{11} \ra = \varphi_1 \,\unit_{N_m\times N_m} \, ,
\qquad \la \phi_{22} \ra = \varphi_2 \,\unit_{N_e\times
N_e} \label{back} \, ,
\ee
with zero vevs for all remaining fields and with $Q$, $\varphi_1$,
$\varphi_2$ all being real.

\subsection{F-terms}

In this subsection we consider the contribution from the F-terms.
The D-terms will be taken into account in the next one.
For convenience, from now on we denote collectively all components
of the fields $q$, $\tilde{q}$ and $\Phi$ simply by $\chi^I$.

\subsubsection{Preliminaries}

Recall that the classical F-term supergravity potential is:\footnote{In
this section we set $M_P=1$. The explicit dependance on the Planck
mass will be reinserted later when needed.}
\be \label{pot}
V = e^K \{ K^{I\bar{J}} D_I W
D_{\bar{J}} \overline{W} - 3 |W|^2 \} \, ,
\ee
where $I,J$ run over all scalar fields in the theory, $K^{I\bar{J}}$ is
the inverse of the K\"{a}hler metric and the K\"{a}hler covariant
derivative is:
\be
D_IW=\partial_I W+\partial_I K \,W \, .
\ee

The supergravity Lagrangian is invariant under K\"{a}hler
transformations:
\bea
K(\chi^I,\bar{\chi}^I)&\rightarrow&
K(\chi^I,\bar{\chi}^I)+F(\chi^I)+\bar{F}
(\bar{\chi}^I) \, , \nonumber\\
W(\chi^I)&\rightarrow& e^{-F(\chi^I)}\, W(\chi^I) \, .
\eea
One can exploit this invariance, by taking $F(\chi^I)=\log W(\chi^I)$,
in order to show \cite{CFGvP} that the scalar potential depends
only on the combination
\be \label{fG}
G = K + {\rm ln} |W|^2  \, ,
\ee
but not on $W$ and $K$ separately. In terms of this function, we
can rewrite (\ref{pot}) as follows:
\be \label{potG}
V =  e^G \left( G^{I} G_I - 3 \right) \,
\ee
where
\be
G_I \equiv \frac{\partial G}{\partial \chi^I} \, , \qquad G^{J}
\equiv \frac{\partial G}{\partial \bar{\chi}_{J}}\,.
\ee
This notation utilizes the fact that for us the K\"{a}hler potential
is canonical i.e. $K_{I\bar{J}} = \delta_{I\bar{J}}$, see (\ref{KaP}),
and so $\bar{\chi}_{J} \equiv K_{J\bar{L}} \,\bar{\chi}^{\bar{L}}$.

For such a K\"{a}hler potential, the expressions for the scalar and
fermionic mass matrices are \cite{BG1}:\footnote{We will consider a
non-canonical K\"{a}hler potential in Section \ref{KKLTISS}, when we
address the KKLT set-up with ISS uplifting sector.}
\be \label{Mb}
{\rm Tr} M_s^2 = \la \,2  e^G \{  (G^{I J} + G^{I} G^{J})
(G_{IJ} + G_I G_J) + (N_B-1) G^{I} G_I - 2 N_B
\} \,\ra \ee
and
\be \label{Mf}
{\rm Tr} M_f^2 ={\rm Tr} M_{1/2}^2+{\rm Tr} M_{3/2}^2
= \la \,e^G \{(G^{I J} + G^{I} G^{J}) (G_{IJ} + G_I G_J) -2
\} \,\ra \, ,
\ee
respectively. Here $N_B$ is the number of complex scalars. Using
(\ref{potG}), it is easy to show that (\ref{Mb}) follows from
\be \label{MbV}
{\rm Tr} M_s^2 = 2 \frac{\partial^2 V}{\partial \chi^J \partial
\bar{\chi}_{J}} \, .
\ee
The derivation of the fermion mass-squared is a bit more involved
since one has to disentangle the mixing between the gravitino and
the goldstino in the supergravity Lagrangian. After having dealt
with that, one finds ${\rm Tr} M_{3/2}^2=-2 e^G$. The anomalous
negative sign in the gravitino contribution is due to the fact
that in ${\rm Tr} M_{1/2}^2$ we have summed over the physical
matter fermions and the goldstino.

Before starting the actual computations, two remarks are in order.
First, it may seem that it is more illuminating to perform the
calculations as in \cite{FKKMT}, i.e. to compute separately the
mass-squareds of every field and then add them. However, for us this
becomes rather cumbersome, whereas the trace formulae above provide
a very efficient way of handling things. And second, the formulation
of the supergravity Lagrangian in terms of the function $G$,
(\ref{fG}), appears to encounter a problem for vanishing
superpotential, as $W$ enters various terms in the denominator.
That will be an issue for the D-terms in Subsection \ref{dterm} and
we will use there a more modern formulation that is equally valid
for $W\neq0$ and $W=0$. Here we simply note that for the F-terms
there is no problem, since the apparent negative powers of $W$,
coming from derivatives of $G$, are cancelled by the positive powers
from $e^G$. (This will be made more explicit in Subsection \ref{MM}.)
So the F-term results never contain division by zero.
This is an important point as in later sections we will be
interested in the effective potential at the origin of field space,
where the ISS superpotential vanishes.

\subsubsection{Mass matrices} \label{MM}

We are finally ready to find the F-term mass matrices
${\rm Tr} M_s^2$ and ${\rm Tr} M_f^2$ for the ISS model
coupled to supergravity. To do so more efficiently, we note
that, instead of finding separate expressions for
the various ingredients $G^{IJ}G_{IJ}$, $G^{IJ}G_I G_J$ and $G^I
G^J G_I G_J$ which enter the mass formulae, it is computationally
much more convenient to calculate the whole combination
$(G^{I J}+G^{I}G^{J})(G_{IJ}+ G_I G_J)$ for each particular choice
of $I$ and $J$. This avoids introducing a big number of terms
that have to cancel at the end, as we now explain. From the
definition of $G$, i.e. $G= K + \ln W + \ln \overline{W}$, we have
that
\be \label{GG}
G_I = K_I + \frac{W_I}{W}, \qquad G_{IJ} =
-\frac{W_I W_J}{W^2} + \frac{W_{IJ}}{W} \, .
\ee
Therefore, in the
expressions $G^{IJ}G_{IJ}$, $G^{IJ}G_I G_J$ and $G^I G^J G_I G_J$
one seems to obtain many terms proportional to $1/ W^2
\overline{W}^2$. Taking into account the $|W|^2$ factor coming
from $e^G$, one is left with many terms $\sim$ $1/|W|^2$. However,
it is clear that they have to cancel at the end, since the scalar
potential (\ref{pot}) does not include any negative powers of
$|W|^2$ and so if we were computing the masses of each field
separately and adding them (as in \cite{FKKMT}) we could not
possibly obtain terms $\sim$ $1/|W|^2$. This cancellation can be
incorporated from the start by using (\ref{GG}) to write:
\be \label{simp0}
G_{IJ} + G_I G_J = \frac{W_{IJ}}{W} + \frac{K_I W_J + K_J W_I}{W}
+ K_I K_J \ee
or equivalently
\be \label{simp}
G_{IJ}+G_IG_J = \frac{W_{IJ}}{W} + G_I K_J + K_I G_J - K_I K_J \, .
\ee
It is evident now that this expression does not contain any
$1/W^2$ terms. Incidentally, this also makes it obvious that,
as expected, the expression $e^G (G^{I J}+G^{I}G^{J})
(G_{IJ}+ G_I G_J)$ does not contain any powers of $W$ in the
denominator.

To illustrate how much the use of (\ref{simp0}) or (\ref{simp})
simplifies the computation, let us look for instance at the
following term:
\be
R^{\phi_{11}\phi_{22}} \equiv (G^{\phi_{11}\phi_{22}}
+G^{\phi_{11}}G^{\phi_{22}})(G_{\phi_{11}\phi_{22}} +
G_{\phi_{11}}G_{\phi_{22}}) \, .
\ee
To make use of (\ref{simp}), we note that:
\be
W_{\phi_{11}\,\phi_{22}} = 0 \, , \qquad \la K_{\phi_{11}}
\ra = \varphi_1 \, \unit_{N_m\times N_m} \, , \qquad \la
K_{\phi_{22}} \ra = \varphi_2 \, \unit_{N_e\times N_e} \, , \nn
\ee
\be
\la G_{(\phi_{11})^i_{\bar{j}}} \ra = \left[ \varphi_1 +
\frac{h}{{W}_0} (Q^2 - \mu^2) \right] \delta^{\bar{j}}_i \, , \qquad
\la G_{(\phi_{22})^k_{\bar{l}}} \ra = \left[ \varphi_2 - \frac{h
\mu^2}{{W}_0} \right] \delta^{\bar{l}}_k \, .
\ee
Therefore, one immediately finds:
\be
\la R^{\phi_{11}\phi_{22}} \ra = \left[
\left( \varphi_1 + \frac{h (Q^2 - \mu^2)}{{W}_0} \right) \varphi_2 +
\left( \varphi_2 - \frac{h \mu^2}{{W}_0} \right) \varphi_1 -
\varphi_1 \varphi_2 \right]^2 (\delta^{\bar{j}}_i
\delta^{\bar{l}}_k) (\delta^i_{\bar{j}} \delta^k_{\bar{l}}) \,,
\ee
where we denoted by $ W_0$  the value of the ISS potential in the
background (\ref{back}). The last factor
gives $(\delta^{\bar{j}}_i \delta^{\bar{l}}_k) (\delta^i_{\bar{j}}
\delta^k_{\bar{l}}) = \delta^i_i \delta^k_k = N_m N_e$. So we obtain
\be
\la R^{\phi_{11}\phi_{22}} \ra = \left[ \varphi_1 \varphi_2 +
\frac{h (Q^2 - \mu^2)}{{W}_0} \varphi_2 - \frac{h \mu^2}{W_0}
\varphi_1 \right]^2 N_m N_e \, .
\ee

Similarly, it is very easy to compute:
\bea \label{Rtwo}
\la R^{\phi_{11}\phi_{11}} \ra &=& \varphi_1^2 \left( \varphi_1 + 2 \frac{h (Q^2 - \mu^2)}
{W_0} \right)^2 N_m^2 \, , \nn \\
\la R^{\phi_{22}\phi_{22}} \ra &=& \varphi_2^2 \left( \varphi_2 - 2 \frac{h \mu^2}{W_0}
\right)^2 N_e^2 \, , \nn \\
\la R^{\,q_1\phi_{11}} \ra = \la R^{\,\tilde{q}_1\phi_{11}} \ra &=& \left( \frac{hQ}{W_0}
\right)^2 N_m^2 +Q^2 \left[ \left( \frac{h\varphi_1}{W_0}+1 \right)\varphi_1 +
\frac{h (Q^2 -\mu^2)}{{W}_0} \right]^2 N_m^2 \nn \\
&+& 2 \frac{hQ^2}{{W}_0} \left[ \left( \frac{h\varphi_1}{{W}_0}+1 \right)\varphi_1
+ \frac{h (Q^2 -\mu^2)}{{W}_0} \right] N_m \, ,\nn \\
\la R^{\,q_1\phi_{22}} \ra = \la R^{\,\tilde{q}_1\phi_{22}} \ra &=& Q^2 \left[ \varphi_2
+ \frac{h}{{W}_0} (\varphi_1 \varphi_2 - \mu^2) \right]^2 N_m N_e \, , \nn \\
\la R^{\,q_2\phi_{21}} \ra = \la R^{\,\tilde{q}_2\phi_{12}} \ra &=& \left(
\frac{hQ}{{W}_0} \right)^2 N_m N_e \, , \nn \\
\la R^{\,q_1 q_1} \ra = \la R^{\,\tilde{q}_1 \tilde{q}_1} \ra &=& Q^4 \left( 1+ \frac{2
h}{{W}_0} \varphi_1 \right)^2 N_m^2 \, , \nn \\
\la R^{\,q_1 \tilde{q}_1} \ra &=& \left( \frac{h \varphi_1}{{W}_0} \right)^2 N_m^2 + Q^4
\left( 1+ \frac{2h \varphi_1}{W_0} \right)^2 N_m^2 + \frac{2h\varphi_1 Q^2}{{W}_0} \left(
1 + \frac{2h \varphi_1}{{W}_0} \right) N_m \nn \\
\la R^{\,q_2 \tilde{q}_2} \ra &=& \left( \frac{h \varphi_2}{{W}_0}
\right)^2 N_m N_e \, .
\eea
For all remaining pairs $\la R^{\,\chi^I \chi^J} \ra = 0$.

The last ingredient in (\ref{Mb}), that we need to compute, is
$\la G^I G_I \ra$. The only nonvanishing components are for
$I = \phi_{11}, \phi_{22}, q_1, \tilde{q}_1$ and we find:
\be \label{GIGI}
\la G^I G_I \ra = \left( \varphi_1 + \frac{h (Q^2 -
\mu^2)}{{W}_0} \right)^2 N_m + \left( \varphi_2 - \frac{h
\mu^2}{{W}_0} \right)^2 N_e + 2Q^2 \left( \frac{h\varphi_1}{{W}_0} +
1 \right)^2 N_m \, .
\ee

One can notice that all expressions above depend only on $Q^2$, not
on $Q$ alone. Hence the temperature dependent part of the one-loop
effective potential as a function of $Q$ is of the form\footnote{We
drop from now on the piece that is $\sim T^4$, as it does not
depend on the vevs of the fields and so it does not contribute to
the derivatives of $V_T$, which are the quantities that will be of
interest for us. \label{Drop}}
\be \label{symbVT}
V_T = e^{Q^2} (A Q^6 + B Q^4 + C Q^2 + D)
\ee
for any values of the meson vevs $\varphi_1$ and $\varphi_2$.
Therefore $Q=0$ is always an extremum. In the meson directions things
are not so apparent, as there are odd powers of $\varphi_1$ and
$\varphi_2$. However, one can see that all of them multiply
either a power of $Q^2$ or the first power of $W_0$. Since $W_0$ is
linear in the meson vevs, for $Q=0$ the dependence of $V_T$ on
$\varphi_1$ and $\varphi_2$ is, in fact, at least quadratic (by that
we also mean mixed terms, i.e. with $\varphi_1 \varphi_2$). We will see
in Section \ref{local} that the point $(Q, \varphi_1, \varphi_2)=(0,0,0)$
is a local minimum of $V_T$, as was also the case for vanishing
supergravity interactions \cite{FKKMT}.\footnote{This is no longer
true if one includes the KKLT sector, as we argue in Section
\ref{KKLTISS}.} Before that, however, let us first consider the
D-term contributions to the mass matrices of the various fields.

\subsection{D-term masses}\label{dterm}

The D-terms for Super Yang-Mills coupled to supergravity were
derived for the first time in \cite{CFGvP}. However that
formulation, entirely in terms of the function $G$ of eq.
(\ref{fG}), is not convenient for our purposes since the D-terms
have a singular dependance on the superpotential. For example,
denoting by $g$ the gauge coupling constant and by $T_{\alpha}$
the generators of the gauge group, the D-term scalar potential
was found to be
\be \label{singVd} V_D =
\frac{1}{2} D^{\alpha} D^{\alpha} \, ,
\ee
with
\be\label{singVd2}
D^{\alpha} = g \,G_I\,
T_{\alpha}{}^{\,I}_{\,J}\, \chi^J = \frac{1}{2} g\,
T_{\alpha}{}^{\, I}_{\,J}\, \chi^J \frac{D_I W}{W}\, ,
\ee
where in the
last equality we have substituted $G^I = K^I + W^I/W$. Clearly,
$V_D$ is not well-defined when the superpotential $W$ vanishes.
The same problem, i.e. division by $W$, appears also in the
fermionic mass terms. So if we adopt the formulation given in
\cite{CFGvP}, Tr$M_f^2$ seems to diverge for $W=0$. That is
problematic since we would like to study the effective
potential at the origin of field space, where the ISS
superpotential vanishes.

It has been noted long ago that the above formulation is not
suitable in the presence of a vanishing
superpotential and the latter case has to be studied
separately, without the use of the function $G = K + \ln |W|^2$.
However, a careful study of the supergravity Lagrangian that is
valid both for $W\neq 0$ and $W=0$ was performed only recently (to
the best of our knowledge) in \cite{KKLvP}. What is relevant for
us is that $V_D$ can be written as
\be \label{genVd}
V_D = \frac{g^2}{2} (i \xi_{\alpha}^{I} \pd_I K - 3 i
r_{\alpha})^2 \, ,
\ee
where
$\xi_{\alpha}^I$ are the gauge transformations of the scalar
fields, i.e. $\delta_{\alpha} \chi^I = \xi_{\alpha}^I (\{
\chi^J \})$, and $r_{\alpha}$ are functions determined by
the gauge variations of the superpotential:
\be \label{dW}
\delta_{\alpha} W = \xi_{\alpha}^I \partial_I W = -3 r_{\alpha}
W \, ,
\ee
where the second equality is
required for gauge invariance of the action. These functions also
characterize the gauge-non-invariance of the K\"{a}hler potential:
\be
\delta_{\alpha} K(\chi, \bar{\chi}) = 3 (r_{\alpha}(\chi) +
\bar{r}_{\alpha}(\bar{\chi})) \, .
\ee
In the case of non-vanishing superpotential one can use
(\ref{dW}) to express $r_{\alpha}$ in terms of
$W$ and $\pd_I W$. Substituting the result in (\ref{genVd}), one
finds (\ref{singVd})-(\ref{singVd2}) upon using $\xi_{\alpha}^I =
i \,T_{\alpha}{}^I_J \,\chi^J$. This, indeed, shows that the
formulation of \cite{KKLvP} reduces to the one in
\cite{CFGvP} when $W\neq 0$.

Since the ISS superpotential is gauge invariant, we have
$r_{\alpha} = 0$ and
\be
V_D = \frac{g^2}{2} \,\sum_{\alpha=1}^{N_m^2 -1} \,\left( {\rm Tr}
(q^{\dagger}_1 T_{\alpha} q_1 - \tilde{q}_1 T_{\alpha}
\tilde{q}^{\dagger}_1)\right)^2
\ee
as in
\cite{FKKMT}. We can then borrow the results, found in the global
supersymmetry case, to obtain a $4g^2Q^2 (N_m^2-1)$ contribution
to Tr$M^2_s$. Also, the vector boson mass is the same as in
\cite{FKKMT} and so it gives a $4g^2Q^2 (N_m^2-1)$ contribution to
Tr$M^2_v$.

Let us now come to the fermionic sector. The mass matrices
are \cite{KKLvP}:
\bea
M^{IJ}&=&{\cal{D}}^I {\cal{D}}^J M\label{mixing1}\\
M_{I\alpha}&=&-i\Big[\partial_I {\cal P}_{\alpha}-{1\over 4}({\rm
Re}f)^{-1\, \beta \gamma}{\cal P}_{\beta}\partial_{I} f_{\gamma
\alpha }\Big]\label{mixing}\\
\qquad M_{\alpha \beta}& =& -{1\over 4}\partial_{\bar{I}} f_{\alpha
\beta}\,K^{\bar{I}J}M_{J}\, ,
\eea
where $f_{\alpha \beta}$ are the gauge
kinetic functions, the action of the covariant derivative ${\cal
D}^I$ on $M\equiv e^{K/2}W$ is
\be
{\cal D}^I=\partial^I+{1\over 2}\partial^I K\,,
\ee
$M^I \equiv {\cal D}^I M$ and finally
\be
{\cal P}_{\alpha} = i \xi_{\alpha}^I \pd_I K  \, .
\ee
These expressions
are clearly well defined and non singular even when $W=0$, unlike
the analogous formulae in \cite{CFGvP}. In our case, obviously the
index $I$ runs now only over the quark fields. For a flat K\"{a}hler
potential, one can easily verify from (\ref{mixing1}) that:
\be
M^{IJ}M_{IJ}=e^G\left(G^{IJ}+G^IG^J\right)\left(G_{IJ}+G_IG_J\right),
\ee
so that we recover correctly the contribution from the
matter fermions and the goldstino in (\ref{Mf}). Since for us
$f_{\alpha \beta} = 1/g^2 = const$ for all $\alpha$ and $\beta$,
$M_{\alpha \beta}=0$ while the
contribution to $TrM_f^2$ from the mixing of gaugino and hyperini
can be found from (\ref{mixing}) using ${\cal P}_{\alpha} =
q^{\dagger}_1 T_{\alpha} q_1 - \tilde{q}_1 T_{\alpha}
\tilde{q}^{\dagger}_1$. It reads
\be
\la 2 M_{I\alpha} M^I_{\alpha} \ra = 8 g^2 Q^2 (N_m^2 - 1) \, .
\ee

In the supergravity Lagrangian there is one more mixing between
fermions, which could potentially add a term to Tr$M_f^2$, namely
the mixing between the gravitino and gaugino. In \cite{KKLvP} it
is of the form $\psi {\cal P}_{\alpha} \lambda^{\alpha}$. However,
in our case $\la {\cal P}_{\alpha} \ra = 0$ and so this mixing does
not contribute. To recapitulate, the D-terms give exactly the same
contribution as in the rigid case considered in \cite{FKKMT}.

\section{Expansion in $M_P^{-1}$ and rigid susy limit}
\setcounter{equation}{0}

In Section \ref{VT}, we computed all ingredients of the finite
temperature one-loop effective potential for the ISS model coupled
to supergravity. Now we will make connection with the globally
supersymmetric theory by inserting in the relevant formulae the
explicit dependance on the Planck mass, $M_P$, and expanding in
powers of $M_P^{-1}$. For later use, we will extract the leading
supergravity corrections to the rigid results.

Let us start with the tree-level supergravity potential
\be \label{potMp}
V = M^4_P\, e^G \left(M^2_P G^I G_I - 3\right) + {1\over2}D^aD^a
\ee
with
\be \label{GMP}
G={K\over M^2_P}+\log{|W|^2\over M_P^6} \, .
\ee
As discussed in the previous section, the D-term
contribution is the same as in the rigid limit. It is
easy to see that the expansion of (\ref{potMp}) gives:
\begin{eqnarray} \label{expp}
V=W^IW_I+{1\over2}D^aD^a +{1\over M^{2}_P}\left( K W^IW_I+2\,Re(K_IW^I
W)-3|W|^2 \right)+{\cal O}(M^{-4}_P)\, ,\,\,\,\,\,\,
\end{eqnarray}
where obviously the first two terms are the standard global susy
result. For future use, we note that taking $I= \phi_{11}, \phi_{22}$ in
the last equation gives, to leading order in the supergravity corrections,
the following contribution to the classical F-term potential for the quarks:
\be \label{pot1}
V_F \supset \left(\!h^2 \, {\rm Tr} \,q_1 q^{\dagger}_1 \tilde{q}_1^\dagger
\tilde{q}_1-h^2\mu^2 ({\rm Tr} q_1 \tilde{q}_1 + {\rm Tr}
q_1^{\dagger} \tilde{q}_1^{\dagger})+h^2\mu^4(N_m+N_e)\!\right)
\!\! \left(\!1+\frac{{\rm Tr}q_1 q_1^\dagger+{\rm Tr}\tilde{q}_1
\tilde{q}_1^\dagger}{M_P^2} \right)\!.
\ee

Inserting the $M_P$ dependence in the thermal one-loop potential
$V_T$ yields\footnote{Recall that, as we mentioned in Footnote \ref{Drop},
we drop for brevity the constant $\sim \,T^4$ contribution to
effective potential in ({\ref{Texpansion}}), as it does not affect
our considerations.}:
\be \label{1lMP} V_{T}={T^2\over 24}M^2_P
\,\la e^G\left(3M^4_P\,\sum_{IJ}
R^{IJ}+2(N_B-1)M^2_PG^IG_I-2(2N_B+1)\right)\ra + \la V_D \ra \, , \ee
where we have denoted by $R^{IJ}$ the quantity
$(G^{IJ}+G^IG^J)(G_{IJ}+G_IG_J)$, as before. To obtain the
explicit powers of $M_P$ in $R^{IJ}$, we note that due to
(\ref{GMP}) equation (\ref{simp}) becomes: \be
G_{IJ}+G_IG_J={W_{IJ}\over W}+{G_IK_J\over M^2_P} +{G_JK_I\over
M^2_P}-{K_IK_J\over M^4_P} \, . \ee Now, expanding (\ref{1lMP}) we
find:
\begin{eqnarray}
V_T&=&{T^2\over 24}\la \Big(3 W^{IJ}W_{IJ}+3{K\over M^2_P}W^{IJ}W_{IJ} +
6{|W|^2\over M^2_P} Re\Big[ {W_{IJ}\over W^2}(W^IK^J+K^IW^J)\Big] \nn \\
&+&2{(N_B-1)\over M^2_P}W^IW_I\Big) \ra + \la V_D \ra + {\cal O}(M^{-4}_P)
\, . \label{exp}
\end{eqnarray}
Together, equations (\ref{expp}) and (\ref{exp}) give the general
expression for the zeroth and first orders in the $M_P^{-1}$ expansion
of the one-loop effective potential at finite temperature.

Let us now apply the above formulae for the background
(\ref{back}), but with vanishing meson vevs. To see what (\ref{exp})
leads to, note that the only contributions to the first two terms
come from $\la |W_{q_1\phi_{11}}|^2 \ra = \la
|W_{\tilde{q}_1\phi_{11}}|^2 \ra = h^2Q^2 N_m^2$ and $\la
|W_{q_2\phi_{21}}|^2 \ra = \la |W_{\tilde{q}_2 \phi_{12}}|^2 \ra =
h^2Q^2 N_mN_e$ and that the only nonzero $\la W^I \ra$ are $\la
W^{\phi_{11}} \ra = h (Q^2 - \mu^2) $ and $\la W^{\phi_{22}} \ra =
- h \mu^2$. Hence we obtain:
\bea \label{VTlo}
V_T&=&{T^2\over2}h^2Q^2(N_m^2+N_mN_e) \left(1+2{Q^2 N_m\over
M_P^2}\right) + T^2 Q^2 g^2 (N_m^2 - 1) \\
&+& { T^2 \over M_P^2} \left( h^2 Q^2 (Q^2 - \mu^2) N_m + {1 \over
12} (N_B -1) [h^2 (Q^2 -\mu^2)^2 N_m + h^2 \mu^4 N_e] \right) +
{\cal O}(M^{-4}_P). \nn
\eea
Taking $M_P \rightarrow \infty$, we find the global supersymmetry result
for $V_T$.\footnote{Our result is in agreement with that of \cite{FKKMT},
upon correcting a typo there. Namely, they have overcounted by a factor of two
the number of Weyl fermions (in their notation) $\Psi_{11}^{\phi}$ and
$(\Psi_1^q+\Psi_1^{\tilde{q}}) / \sqrt{2}$.}

\subsection{Local minimum at the origin}\label{local}

At high temperature, the temperature-dependent contribution $V_T$
completely dominates the effective potential and so the minima of
$V_{eff}$ are given by the minima of $V_T$. Let us now address the
question whether the origin of field space is a local minimum of $V_T$.

In principle, this could be a complicated problem, as we have to
consider a function of three variables, $V_T (Q, \varphi_1, \varphi_2)$.
However, things are enormously simplified by the fact that, as we noted
at the end of Subsection \ref{MM}, $Q=0$ is an extremum for any
$\varphi_1$ and $\varphi_2$. Since it is also a local minimum in the
rigid limit, it will clearly remain such when taking into account the
subleading supergravity corrections in our case. So we are left with
investigating a function of two variables, $\varphi_1$ and $\varphi_2$.
The presence of terms linear in any of them could, potentially, shift the
position of the minimum away from the point $(\varphi_1, \varphi_2) =
(0,0)$.\footnote{This is what happens for the KKLT-ISS model, as we
will see in Section \ref{KKLTISS}.} However, such terms in the ISS model
coupled to supergravity appear only multiplied by powers of $Q^2$, as we
noted below eq. (\ref{symbVT}). So, given that $(\varphi_1, \varphi_2) =
(0,0)$ is a local minimum in the global supersymmetry case, it will remain
such after coupling to supergravity, to all orders in the $1/M_P$ expansion.

It is, nevertheless, instructive to write down explicitly the expression
for $V_T$ to leading order in the $1/M_P$ corrections:
\bea \label{VTMP}
V_T (Q, \varphi_1, \varphi_2) &=& \frac{T^2}{2} h^2 Q^2 (N_m^2 + N_e N_m) +
T^2 Q^2 g^2 (N_m^2 - 1) + \frac{T^2}{4} h^2 (\varphi_1^2 N_m^2 +
\varphi_2^2 N_m N_e) \nn \\
&+& \frac{T^2}{M_P^2} \left[ h^2 Q^2 (2 \varphi_1^2 + Q^2 - \mu^2) N_m +
\frac{(N_B-1)}{12} \{ h^2 ((Q^2 - \mu^2)^2 + 2 \varphi_1^2) N_m \right. \nn \\
&+& h^2 \mu^4 N_e \} + (2 Q^2 N_m + \varphi_1^2 N_m + \varphi_2^2 N_e)
 \left(\frac{1}{2} h^2 Q^2 (N_m^2 + N_m N_e) \right. \nn \\
&+& \left. \left. \frac{1}{4} h^2 (\varphi_1^2
N_m^2 + \varphi_2^2 N_m N_e)\right) \right] + {\cal O} \left(\frac{1}{M_P^4}
\right).
\eea
Clearly, this is consistent with (\ref{VTlo}).
We have collected the terms that survive in the $M_P \rightarrow \infty$
limit on the first line. Obviously, the origin of field space $(Q,
\varphi_1, \varphi_2) = (0,0,0)$ is a minimum. One can also notice that to
this order $V_T$ is a function of the form $V_T (Q^2, \varphi_1^2,
\varphi_2^2)$, i.e. it does not depend on odd powers of any of the vevs.
One can verify that the terms linear in $\varphi$ and multiplying powers of
$Q^2$, that we mentioned above, start appearing at ${\cal O} (1/M_P^6)$,
while terms linear in $\varphi_1 \varphi_2$ appear at ${\cal O} (1/M_P^4)$.

It will be useful for the next sections to
extract from (\ref{VTMP}) a couple of special cases. One case is the
second derivative in the quark direction only:
\be \label{VTQ2}
\frac{\pd^2 V_T}{\pd Q^2}\Big|_{Q=0, \varphi_1 = 0, \varphi_2
= 0} = T^2 [h^2 (N_m^2 + N_m N_e) + 2 g^2 (N_m^2 -1)] -
\frac{T^2}{M_P^2} \,\frac{1}{3} h^2 \mu^2 N_m (N_B+5) \, .
\ee
The other interesting case is the potential in the meson
directions only:
\bea \label{Vmesonlo}
V_T (0, \varphi_1, \varphi_2) &=& \frac{T^2}{4} \left[ h^2 \left(
\varphi_1^2 N_m^2 + \varphi_2^2 N_m N_e \right) + \frac{h^2}{M_P^2}
\left( \varphi_1^2 N_m + \varphi_2^2 N_e \right)^2 N_m \right. \nn \\
&+& \left. \frac{1}{M_P^2} \,\frac{1}{3} h^2 \mu^4 (N_B -1) (N_m +
N_e) \right] .
\eea

Finally, let us note that if we want to study other minima of the
potential, which are away from the origin of field space $(Q, \varphi_1,
\varphi_2) = (0,0,0)$, we would have to include in our considerations
the non-perturbative superpotential $W_{ADS}$, see eq. (\ref{WADS}).

\section{Phase transition in quark direction} \label{PhTrQ}
\setcounter{equation}{0}

As we saw in the previous section, at high temperature the origin of
field space is a local minimum of the effective potential.
Lowering the temperature, one reaches a point at which
this minimum becomes unstable and the fields start evolving towards new
vacua. As recalled in Section \ref{ISSrev}, at zero temperature
the ISS model possess supersymmetric vacua at non-zero meson vev
and a metastable vacuum in the quark branch. Adding the supergravity
interactions results in a slight shift of the positions of those minima.
In particular, the metastable vacuum shifts to small vevs for some of
the meson directions \cite{AHKO}. Hence, in order to end up in it,
the system must undergo a second order phase transition in those same
meson directions, as well as in the quark ones. In the next section we
will see that this is indeed the case. In the present section, we will
find the critical temperature, $T_c^Q$, for the onset of a second order
phase transition towards nonvanishing quark vevs, while keeping all
meson vevs at zero.

Before turning to the supergravity corrections to $T_c^Q$, let us
recall how to compute the critical temperature in a generic field
theory \cite{DJ}. Suppose that we have a theory with a set of
fields $\{\chi^I \}$. To find the effective potential, one has to
shift $\chi^I$ by constant background fields $\hat{\chi}^I$, as we
reviewed in Section \ref{VT}. The effective potential is a
function of $\hat{\chi}^I$ only and we consider the case when it
is of the form $V_{eff}(\hat{\chi}^{2})$.\footnote{This will be
the case for us. However, even in principle this assumption is not
a restriction but merely a simplification, as we explain in a
subsequent footnote.} The location of the minima is then
determined by:
\be
{\partial V_{eff}(\hat{\chi}^2)\over\partial
\hat{\chi}^I}=2\hat{\chi}^I {\partial
V_{eff}(\hat{\chi}^2)\over\partial \hat{\chi}^2}=0 \, .
\ee
Clearly, $\hat{\chi}^I = 0$ satisfies this condition. Other
minima, at $\hat{\chi}^I\neq 0$, can only occur when $\partial
V_{eff}/\partial \hat{\chi}^2 = 0$. The critical temperature,
at which rolling towards such minima begins, can be found by
requiring \cite{DJ}:
\be \label{split}
{\partial V_{eff}\over\partial \hat{\chi}^2}\Big|_{\hat{\chi}=0}=
{\partial V_0\over\partial \hat{\chi}^2}\Big|_{\hat{\chi}=0}+
{\partial V_T\over\partial \hat{\chi}^2}\Big|_{\hat{\chi}=0}=0 \, ,
\ee
where we have split the one-loop effective potential into
zero-temperature and temperature-dependant contributions, $V_0$
and $V_T$ respectively. This is equivalent to\footnote{The
assumption that $V_{eff}$ depends only on $\hat{\chi}^2$, but not
on $\hat{\chi}$ alone, guarantees that $\hat{\chi} = 0$ is an
extremum of the effective potential and thus simplifies the
computations technically. In the general case, instead of
(\ref{split1}) one has to solve, symbolically, the following
system: $V^{\prime}_{eff} (\hat{\chi}_c, T_c) = 0 \, , \,V^{\prime
\prime}_{eff} (\hat{\chi}_c, T_c) = 0$ in order to find the
critical temperature $T_c$, at which the phase transition occurs,
{\it and} the vev $\hat{\chi}_c$, at which the relevant extremum
of $V_{eff}$ is situated. Clearly, here we have denoted by $'$ and
$''$ first and second derivatives w.r.t. to $\hat{\chi}$,
respectively.}
\be \label{split1}
{\partial V_T\over\partial \hat{\chi}^2}\Big|_{\hat{\chi}=0}=-
\frac{m^2}{2} \, ,
\ee
where $m^2$ is the unshifted tree-level\footnote{In principle, $m^2$
gets also contributions from the Coleman-Weinberg potential. However,
in our case the origin of field space is not a local minimum of the
zero temperature potential and so perturbation theory around it
does not make sense. Hence, for us, $m^2$ is purely classical. In fact,
to be more precise, we should note that at the origin the Coleman-Weinberg potential becomes imaginary. As shown in \cite{WW}, the imaginary part of
the effective potential encodes the decay rate of a system
with a perturbative instability. This decay process leads to the breaking
up of the initial homogeneous field configuration into various domains, in
each of which the fields are rolling toward a different classical solution,
and should not be confused with a non-perturbatively induced tunneling
between different minima of the potential. We will not dwell on that
further in the present paper.} mass-squared of the field,
whose nonzero vev characterizes the new vacuum. This last equation
can only be solved if $m^2<0$. In the global supersymmetry case,
among the quark fields, which have nonzero vevs at the metastable
minimum, the only ones with negative tree-level mass-squared are
the components of $Re (q_1 \!+ \tilde{q}^{\,t}_1)/ \sqrt{2} \equiv Re
q_+$ \,\cite{FKKMT}. Since for small vevs the gravitational
corrections are subleading, we are guaranteed that this will be
the case for us as well.

\subsection{Mass matrix diagonalization} \label{Diag}

We will now compute the tree-level squared masses of the fields $Re q_+$, that
are necessary for finding the critical temperature in the
quark-direction. Before considering the supergravity corrections, it will be
useful to give the derivation of these masses in the global
supersymmetry limit. For more generality, we keep $Q \neq 0$ in this latter case,
although we will take $Q=0$ when we turn to the supergravity contributions to
$m^2$ in (\ref{split1}).

From the tree-level scalar potential of the global theory, $V =
K^{I\bar{J}} \pd_I W \pd_{\bar{J}} \bar{W}$, one finds \cite{FKKMT}:
\be \label{glob_nonD}
h^2 Q^2 (|q_1|^2 + |\tilde{q}_1|^2) + h^2 Q^2 \,
{\rm Tr} (q_1 \tilde{q}^{\dagger}_1 + q^{\dagger}_1 \tilde{q}_1) + h^2
(Q^2 - \mu^2) ({\rm Tr} \tilde{q}_1 q_1 + h.c.) \, .
\ee
When we turn on gravitational
interactions we will have a similar expression but, generically, with
different coefficients. So it is of benefit to consider the general
expression:
\be \label{nonDiag}
A (|q_1|^2 + |\tilde{q}_1|^2) + B ({\rm Tr} q_1 \tilde{q}^{\dagger}_1 +
{\rm Tr} q^{\dagger}_1 \tilde{q}_1) + C ({\rm Tr} \tilde{q}_1 q_1 + h.c.)
\ee
with arbitrary $A$, $B$ and $C$. It can be diagonalized easily by
introducing the combinations
\be \label{qpqm}
q_+ = \frac{q_1 + \tilde{q}^{\,t}_1}{\sqrt{2}} \, , \qquad q_- = \frac{q_1
- \tilde{q}^{\,t}_1}{\sqrt{2}} \, .
\ee
Substituting the inverse transformation,
\be
q_1 = \frac{q_+ + q_-}{\sqrt{2}} \qquad {\rm and} \qquad \tilde{q}^{\,t}_1 =
\frac{q_+ - q_-}{\sqrt{2}} \, ,
\ee
in (\ref{nonDiag}), we find:
\be
A ( |q_+|^2 + |q_-|^2 ) + B (|q_+|^2 - |q_-|^2) + C \,\frac{1}{2} \,{\rm Tr}(q_+^2
- q_-^2 + q^{\dagger 2}_+ - q^{\dagger 2}_-) \, .
\ee
Finally, we decompose $q_{\pm} = Re (q_{\pm})
+ i Im (q_{\pm})$ and obtain the following mass terms:
\be \label{MMdiag}
(A + B + C) (Re (q_+))^2 + (A + B - C) (Im (q_+))^2 +
(A - B - C) (Re (q_-))^2 + (A - B + C) (Im (q_-))^2 \, .
\ee

Reading off the values of $A$, $B$ and $C$ from (\ref{glob_nonD}),
we obtain from (\ref{MMdiag}):
\bea
m^2_{Re q_+} = h^2 (3Q^2 - \mu^2) \, &,& \,\, m^2_{Im q_+} = h^2 (Q^2 +
\mu^2) \, , \nn \\
m^2_{Re q_-} = h^2 (\mu^2 - Q^2) \, &,& \,\, m^2_{Im q_-} = h^2 (Q^2
- \mu^2) \, . \label{masse}
\eea
We see that, as already mentioned above, only the components $Re
q_+$ of the field $q_+$ have negative mass-squareds for zero shift
$Q$.\footnote{Clearly, the negative mass-squared of $Im q_-$ is
irrelevant as in the metastable supersymmetry breaking vacuum
$\la q_- \ra = 0$.} Therefore, only their masses should appear on the
right hand side of eq. (\ref{split1}). We note that the masses for
the fields $Im q_+$ and $Im q_-$ differ from those given in table 5
of \cite{FKKMT}. Fortunately their typos cancel out in the total
Tr$M^2_s$.

Let us now apply the result (\ref{MMdiag}) for the diagonalization
of the  expression (\ref{nonDiag}) to the case of interest for us.
Namely, we consider the ISS model coupled to supergravity and we
want to compute the following derivatives:
\be
\pd^2_{q_1 q^{\dagger}_1}
V \, , \,\, \pd^2_{q_1 \tilde{q}^{\dagger}_1} V \, , \,\,
\pd^2_{\tilde{q}_1 {q}_1} V \,\, {\rm etc.} \, ,
\ee
which give the coefficients $A$,
$B$ and $C$. As before, $V$ is the tree-level supergravity scalar
potential. The only nonvanishing derivatives, upon setting $Q=0$,
are:\footnote{For more details see Appendix A. Note that
(\ref{mixedMM}) are exact expressions, i.e. to all orders in
$1/M_P$. However, since they happen to be at most of ${\cal
O}(\frac{1}{M_P^2})$, they can also be read off from the part of the
scalar potential (\ref{pot1}), which contains only the relevant leading
order supergravity corrections.}
\be \label{mixedMM}
m^2_{q_1 q^{\dagger}_1} = m^2_{\tilde{q}_1 \tilde{q}^{\dagger}_1} = \frac{1}{M_P^2} h^2
\mu^4 (N_m+N_e) \, , \qquad m^2_{q_1 \tilde{q}_1} = m^2_{q^{\dagger}_1
\tilde{q}^{\dagger}_1} = - h^2 \mu^2 \, .
\ee
Hence, applying
(\ref{MMdiag}), we find that the tree-level mass of each of the
$N_m^2$ real fields in $Re q_+$ is:
\be \label{mqplus}
m^2_{Re q_+} = -h^2 \mu^2 +\frac{1}{M_P^2} h^2 \mu^4 (N_m + N_e) \, .
\ee
The first term corresponds to the global supersymmetry result, as
can be seen from (\ref{masse}), while the second is a correction due
to the supergravity interactions.

Note that we did not need to include the D-terms in the computation
of the unshifted masses. The reason is that, as we have seen in Section
\ref{dterm}, the contribution of these terms to the mass-squared of the
fields is always proportional to $Q$ and so vanishes for zero shift.

\subsection{Critical temperature} \label{QuTc}

To find the critical temperature in the quark direction, recall that
we shift the fields $q_1{}^a_i$ and $\tilde{q}_1{}^{\bar{j}}_b$ by
constant matrices $Q \delta^a_i$ and $Q \delta^{\bar{j}}_b$
respectively, see eq. (\ref{back}). In the basis of $q_+$ and $q_-$,
see equation (\ref{qpqm}), the only fields which get shifted are the
$N_m$ diagonal components of $q_+$. From those, only the $N_m$ fields
${\rm Re} q_+$ have negative tree-level mass-squared as we saw above.
Therefore we can find the critical temperature
from\footnote{Clearly, this is equivalent to (\ref{split1}) as
$\pd^2 V_T/ \pd Q^2 = 2 \,\pd V_T / \pd (Q^2)$.}
\be
{\partial^2 V_T\over\partial {Q^2}}\Big|_{Q=0}=-N_m \,m^2_{{\rm Re} q_+}
\label{split2}\, .
\ee
Recall that, to leading order in the supergravity corrections, we have
(see (\ref{VTQ2})):
\begin{eqnarray}
\frac{\pd^2 V_T}{\pd Q^2} \Big|_{Q=0} = T^2 \left[
h^2 (N_m^2 + N_m N_e) + 2 g^2 (N_m^2 - 1) \right] -
\frac{1}{M_P^2} \frac{T^2 h^2 \mu^2 N_m (N_B + 5)}{3} \, .
\end{eqnarray}
Together with (\ref{mqplus}), this then leads to:
\bea \label{TcQ}
(T_c^Q)^2 &=& \frac{h^2 \mu^2 N_m}{ h^2 (N_m^2 + N_m N_e) + 2 g^2
(N_m^2 - 1) } - \frac{1}{M_P^2} \,\frac{ h^2 \mu^4 (N_m + N_e) N_m}{h^2
(N_m^2 + N_m N_e) + 2 g^2 (N_m^2 - 1)} \nn \\
&+& \frac{1}{M_P^2} \,\frac{ h^4 \mu^4 N_m^2 (N_B + 5)}{3\,[h^2
(N_m^2 + N_m N_e) + 2 g^2 (N_m^2 - 1)]^2} + {\cal O} \left(\frac{1}{M_P^4}\right) \, .
\eea

To see whether the order $1/M_P^2$ correction increases or decreases the
critical temperature, recall that $N_f > 3 N$ i.e. $N_e > 2 N_m$. Let us write
\be
N_e = 2 N_m + p \, , \qquad p \in \mathbb{Z}_+
\ee
and substitute this in the total numerator of the ${\cal O}(\frac{1}{M_P^2})$
terms in (\ref{TcQ}):
\be
Numerator = -h^4 \mu^4 \!\left(\!4N_m^2 + \frac{10}{3} N_m p + \frac{2p^2 - 5}{3}
\right) \!N_m^2 - 2 h^2 \mu^4 g^2 (3N_m^2 + p N_m) (N_m^2 -1) \, .
\ee
Clearly, for any $N_m$ and for any $p>1$ every term in the above expression is
negative definite. For $p=1$ there is a positive contribution from the last term
in the first bracket: $h^4 \mu^4 N_m^2$. However, it is outweighed by the first
two terms in that bracket for any value of $N_m$. So the conclusion is that the
supergravity interactions cause $T_c^Q$ to decrease compared to the rigid case
for any $N_m$ and $N_e$.

\section{Phase transition in meson direction} \label{mesonPhtr}
\setcounter{equation}{0}

In the global supersymmetry case, a second order phase transition is only possible
in a field space direction with nonzero squark vevs \cite{FKKMT}.\footnote{Although
a first order phase transition can still occur in the meson direction \cite{FKKMT}.}
However, in our case things may be quite different due to the supergravity-induced
contributions to the tree-level meson masses. Note that, in the approximation of
neglecting $W_{dyn}$, the meson fields have zero classical mass in the global limit
\cite{FKKMT} and so the supergravity corrections are the leading ones.\footnote{Recall
that we do not have to take into account contributions from zero-temperature one-loop
effects in the rigid theory as the origin of field space is not a local minimum of the
zero-temperature potential.} Thus the possibility occurs that in the region of small
vevs, for which neglecting the dynamically generated superpotential is well-justified,
some of the meson directions may develop negative tree-level mass-squareds due to
supergravity. We will see in the following when that happens.

Let us address this issue in a slightly more general set-up as in \cite{AHKO}, namely by
adding to the ISS superpotential a constant piece. I.e., we consider $W=W_0 + W_{ISS}$,
where $W_{ISS}$ is as in (\ref{Wtree}) and $W_0=const$. This is also a useful preparation
for the KKLT-ISS model that we will discuss more in the next subsection.
It will turn out that we need to compute the tree-level supergravity-induced meson
masses not only at the origin of field space but also along the quark direction, i.e. for
$Q\neq 0$ and $\varphi_1, \varphi_2 = 0$.\footnote{For reasons that will become clear
below, we will be interested only in the range $Q \in [0, \mu]$. So we are still allowed
to neglect $W_{dyn}$.} Also, we will have to keep track of terms of up to ${\cal O}
(1/M_P^4)$ to see the novel effect.

Turning to the computation, from (\ref{ApMM}) we find:
\bea \label{MMQ}
m^2_{(\phi_{11})^i_{\bar{j}} (\bar{\phi}_{11})^{\bar{l}}_k} &=& 2 h^2 Q^2
\delta_i^{\bar{j}} \delta_{\bar{l}}^k + \frac{1}{M_P^2} \left\{ \delta^k_i
\delta^{\bar{j}}_{\bar{l}} \, h^2 \!\left[ (Q^2 - \mu^2 )^2 N_m + \mu^4 N_e \right] \right.
\nn \\
&+& \left. \delta^{\bar{j}}_i \delta^k_{\bar{l}} \, h^2 \!\left[ (Q^2 - \mu^2) (3 Q^2 +
\mu^2) + 4 Q^4 N_m \right] \right\} \nn \\
&+& \frac{1}{M_P^4} \left\{ \delta^k_i \delta^{\bar{j}}_{\bar{l}} \left[ -2 W_0^2 +
2 h^2 Q^2 N_m \left( (Q^2 - \mu^2)^2 N_m + \mu^4 N_e \right) \right] \right. \nn \\
&+& \left. \delta^{\bar{j}}_i \delta^k_{\bar{l}} \, 4 h^2 Q^4 N_m \!\left[2 (Q^2 - \mu^2)
+ Q^2 N_m \right] \right\} \,\, , \nn \\
m^2_{(\phi_{22})^m_{\bar{n}} (\bar{\phi}_{22})^{\bar{p}}_q} &=& \frac{1}{M_P^2} \left\{
\delta^q_m \delta^{\bar{n}}_{\bar{p}} \, h^2 \!\left[ (Q^2 - \mu^2)^2 N_m + \mu^4 N_e
\right] - h^2 \mu^4 \delta^{\bar{n}}_m \delta^q_{\bar{p}} \right\} \nn \\
&+& \frac{1}{M_P^4} \,\, \delta^q_m \delta^{\bar{n}}_{\bar{p}} \left\{ -2 W_0^2 +
2 h^2 Q^2 N_m \left[ (Q^2 - \mu^2)^2 N_m + \mu^4 N_e \right] \right\} \,\, , \nn \\
m^2_{(\phi_{11})^i_{\bar{j}} (\bar{\phi}_{22})^{\bar{p}}_q} &=& - \,h^2 \mu^2 \left\{
\frac{ Q^2 \!+ \mu^2}{M_P^2} + \frac{4 Q^4 N_m}{M_P^4} \right\} \delta^{\bar{j}}_i
\delta^q_{\bar{p}} \,\,\, , \nn \\
m^2_{(\phi_{22})^n_{\bar{m}} (\bar{\phi}_{11})^{\bar{l}}_k} &=& - \, h^2 \mu^2 \left\{
\frac{Q^2 \!+ \mu^2}{M_P^2} + \frac{4 Q^4 N_m}{M_P^4} \right\} \delta^{\bar{n}}_m
\delta^k_{\bar{l}} \,\, ,
\eea
where clearly \,$i,j,k,l = 1, ... , N_m$ \,and \,$m,n,p,q = 1,..., N_e$. Note that $W_0$
appears only at order $1/M_P^4$. All terms of the form $m^2_{\phi_{11} \phi_{11}}$,
$m^2_{\phi_{22} \phi_{22}}$ and $m^2_{\phi_{11} \phi_{22}}$
vanish (see Apendix A). To understand which
mass matrix one has to diagonalize, let us recall that the meson
fields are shifted as follows: $(\phi_{11})^i_{\bar{j}}$ by
$\varphi_1 \delta^i_{\bar{j}}$ and $(\phi_{22})^m_{\bar{n}}$ by
$\varphi_2 \delta^m_{\bar{n}}$, see eq. (\ref{back}). Hence there are $N_m$ fields
shifted by $\varphi_1$ (let us denote them by $\phi_1$) and $N_e$
fields shifted by $\varphi_2$ (let us denote them by $\phi_2$). In
the quark direction we could factor out an overall $N_m$ (since
both $q_1$ and $\tilde{q}_1$ have the same number of components)
and diagonalize the remaining mass matrix.\footnote{This is why we
ended up with $m^2 = N_m m^2_{Re q_+}$ in (\ref{split2}), where
$m^2_{Re q_+}$ was the result of the diagonalization.} Now things
are somewhat more complicated as there are different numbers of
$\phi_1$ and $\phi_2$ fields. More precisely, from (\ref{MMQ})
we see that we have to diagonalize the expression:
\bea \label{MpQ}
M_{\phi}^2 &=& \left( 2h^2Q^2 N_m^2 + \frac{1}{M_P^2} (N_m C_1 + N_m^2 C_2) +
\frac{1}{M_P^4} (N_m C_3 + N_m^2 C_4) \right) \phi_1 \bar{\phi}_1 \nn \\
&+& \left( \frac{1}{M_P^2} \, h^2 \!\left( Q^2 - \mu^2 \right)^2 N_m N_e +
\frac{1}{M_P^4} N_e C_3 \right) \phi_2 \bar{\phi}_2 \nn \\
&-& h^2 \mu^2 \,N_m N_e \left( \frac{Q^2+\mu^2}{M_P^2} + \frac{4Q^4 N_m}{M_P^4} \right)
\left( \phi_1 \bar{\phi}_2 + \bar{\phi}_2 \phi_1 \right) \, ,
\eea
where
\bea \label{C1C4}
C_1 &=& h^2 \left[ \left(Q^2 - \mu^2 \right)^2 N_m + \mu^4 N_e \right] \, , \nn \\
C_2 &=& h^2 \left[ (Q^2 - \mu^2) (3 Q^2 + \mu^2) + 4 Q^4 N_m \right] \, , \nn \\
C_3 &=& -2 W_0^2 + 2 \,Q^2 N_m \,C_1 \, ,
\nn \\
C_4 &=& 4 h^2 Q^4 N_m \left( 2 (Q^2 - \mu^2) + Q^2 N_m \right) \, .
\eea
Notice that $W_0$ enters the above formulae only at order $1/M_P^4$.

Before proceeding further, let us make an important remark about the position of the local minimum that is our starting point at high temperature. From (\ref{Rtwo}) and (\ref{GIGI}), one can see that for $W_0 \neq 0$ terms linear in $\varphi_1$ and $\varphi_2$ start appearing at order $1/M_P^4$. This implies that the position of the minimum is shifted to some point $(Q,\varphi_1,\varphi_2) = (0, \varphi_1^*, \varphi_2^*)$ with $\varphi_1^*$, $\varphi_2^* \neq 0$. However, one can easily verify that both $\varphi_1^*$ and $\varphi_2^*$ are of ${\cal O}(1/M_P^4)$.\footnote{More precisely, one finds $\varphi_1^* = \frac{4 \mu^2 W_0}{h N_m} \frac{1}{M_P^4} + {\cal O}(\frac{1}{M_P^6})$ and $\varphi_2^* = \frac{4 \mu^2 W_0}{h N_m} \frac{1}{M_P^4} + {\cal O}(\frac{1}{M_P^6})$. We have not looked at  the subleading orders, so we abstain from claiming that $\varphi_1^* = \varphi_2^*$.} Therefore, in all terms of (\ref{Ap1}) with an explicit $1/M_P$ dependence one can take $\varphi_1^*$, $\varphi_2^* = 0$ since we are working to order $1/M_P^4$. Hence the only place the nonzero $\varphi_1^*$, $\varphi_2^*$ can make a difference in is the zeroth order terms. This translates to the zeroth order term in (\ref{ApMM}). However, that term is independent of the meson vevs as the superpotential is linear in those and the indices $I,J$ are along meson directions. Thus the results (\ref{MMQ})-(\ref{C1C4}) are unchanged by the nonzero $\varphi_1^*$, $\varphi_2^*$. For convenience, in the following we will keep refering to the local minimum at $(0,\varphi_1^*,\varphi_2^*)$ as 'the minimum at the origin of field space'.

Let us now go back to (\ref{MpQ}). Since the coefficients of $\phi_1 \bar{\phi}_1$ and $\phi_2 \bar{\phi}_2$ are not the
same, we cannot diagonalize this expression immediately by using the results of
Subsection \ref{Diag}. However, it is still useful to change basis to the real
components of the fields: $\phi_1 = Re \phi_1 + i Im \phi_1$ and
$\phi_2 = Re \phi_2 + i Im \phi_2$. Then $M_{\phi}^2$ acquires the form:
\be \label{MatrixM}
M_{\phi}^2 = \sum_{i=1}^2 (M_{11} \,x_i^2 + 2 M_{12} \,x_i y_i + M_{22} \,y_i^2) \, ,
\ee
where $(x_1,y_1) = (Re \phi_1, Re \phi_2)$ and $(x_2,y_2) = (Im \phi_2, Im \phi_2)$.
Thus the problem is reduced to the diagonalization of the $2\times 2$ matrix $M_{ab}$
with $a,b = 1,2$ and so the eigenvalues are given by:
\be \label{TrDet}
m_{\pm}^2 = \frac{ {\rm Tr} M \pm \sqrt{({\rm Tr} M)^2 - 4 \,{\rm det} M} }{2} \, .
\ee
Before proceeding further with the $Q\neq 0$ considerations, let
us first take a look at what happens at the point $(0,\varphi_1^*,\varphi_2^*)$, which for us, as explained above, is the same as looking at the origin of field space.

For $Q=0$, the expressions (\ref{MMQ}) and (\ref{MpQ}) simplify significantly. Namely, we
have:
\bea \label{MnondM}
m^2_{(\phi_{11})^i_{\bar{j}} (\bar{\phi}_{11})^{\bar{l}}_k}\big|_{Or} &=& \frac{h^2
\mu^4}{M_P^2} \left[ \delta^k_i \delta^{\bar{j}}_{\bar{l}} (N_m + N_e) -
\delta^{\bar{j}}_i \delta^k_{\bar{l}} \right] - \frac{2 W_0^2}{M_P^4} \,\delta^k_i
\delta^{\bar{j}}_{\bar{l}} \,\,\, , \nn \\
m^2_{(\phi_{22})^m_{\bar{n}} (\bar{\phi}_{22})^{\bar{p}}_q}\big|_{Or} &=& \frac{h^2
\mu^4}{M_P^2} \left[ \delta^q_m \delta^{\bar{n}}_{\bar{p}} (N_m + N_e) -
\delta^{\bar{n}}_m \delta^q_{\bar{p}} \right] - \frac{2 W_0^2}{M_P^4} \,\delta^q_m
\delta^{\bar{n}}_{\bar{p}} \,\,\, , \nn \\
m^2_{(\phi_{11})^i_{\bar{j}} (\bar{\phi}_{22})^{\bar{p}}_q}\big|_{Or} &=& - \,\frac{h^2
\mu^4}{M_P^2} \,\delta^{\bar{j}}_i \delta^q_{\bar{p}} \,\,\,\, , \quad \,
m^2_{(\phi_{22})^n_{\bar{m}} (\bar{\phi}_{11})^{\bar{l}}_k}\big|_{Or} = -
\,\frac{h^2 \mu^4}{M_P^2} \,\delta^{\bar{n}}_m \delta^k_{\bar{l}} \,\,\, ,
\eea
where $\big|_{Or}$ denotes evaluation at the origin, and therefore:
\bea \label{Mphif}
M_{\phi}^2\big|_{Or} \!&=& \!\left[ \frac{h^2 \mu^4}{M_P^2} \, N_m N_e - \frac{2 W_0^2}{M_P^4} N_m \right] \!\phi_1 \bar{\phi}_1 + \left[ \frac{h^2 \mu^4}{M_P^2} \, N_m N_e - \frac{2 W_0^2}{M_P^4} N_e \right] \!\phi_2 \bar{\phi}_2 \nn \\
\!&-& \!\frac{h^2 \mu^4}{M_P^2} \, N_m N_e \!\left(\phi_1 \bar{\phi}_2 + \bar{\phi}_1 \phi_2 \right) \!.
\eea
Applying (\ref{TrDet}), we find that the two eigenvalues are:
\be \label{EigO}
m_+^2 = \frac{2 h^2 \mu^4}{M_P^2} N_m N_e - \frac{W_0^2}{M_P^4} (N_m + N_e) \, , \qquad m_-^2 = - \frac{W_0^2}{M_P^4} (N_m + N_e) \, .
\ee
So, clearly, for any $N_m$ and $N_e$ there is a negative meson mass-squared direction.
However, note that its value is of lower order in $1/M_P$ than the leading term (which is
of zeroth order) in the negative squark mass-squared that is driving the quark phase
transition; see (\ref{mqplus}). Hence even without calculating the critical temperature
$T_c^{\varphi}$, that would correspond to $m_-^2$, we can be sure that it would be much
smaller than $T_c^Q$ of Subection \ref{QuTc}. Therefore, by the time the temperature
starts approaching $T_c^{\varphi}$, the system would have already undergone the second
order phase transition in the quark direction and would be rolling along the $Q$ axis. So
let us get back to considering (\ref{TrDet}) with $Q\neq 0$.

One easily finds that, up to order $1/ M_P^4$, the two eigenvalues of the matrix $M$ are:
\bea \label{EigD}
m^2_+ &=& 2 h^2 Q^2 N_m^2 + \,{\cal O} \!\left( \frac{1}{M_P^2} \right) \, , \nn \\
m^2_- &=& \frac{h^2 (Q^2 - \mu ^2)^2 N_m N_e}{M_P^2} + \frac{1}{M_P^4}
\,\frac{N_e}{2 \,Q^2} \left[ -4 Q^2 W_0^2 + 4 h^2 Q^4 (Q^2 - \mu^2)^2 N_m^2 \right. \nn \\
&+& \left. 4 h^2 \mu^4 Q^4 N_m N_e - h^2 \mu^4 (Q^2 + \mu^2)^2 N_e \right] \,.
\eea
The expression for $m_-^2$ seems divergent for $Q=0$. However, we just saw in the previous
paragraph that at the point $Q=0$ things are completely regular.\footnote{This is, in
fact, the main reason we performed that computation explicitly; it was already clear from (\ref{MMQ}) that if there is a critical temperature $T_c^{\varphi}$ at $Q=0$, then it must be that $T_c^{\varphi}<\!\!< T_c^Q$.} The reason for the apparent problem in (\ref{EigD}) is the following. To
obtain the last formulae, we had to expand the square root in (\ref{TrDet}) for small
$1/M_P$. This is perfectly fine when $Q$ is of order $\mu$. However, when $Q\rightarrow
0$ one has to be more careful as the "leading" zeroth order contribution, $2 h^2 Q^2$, in
the above mass formulae becomes $<\!\!<\mu^4/M_P^2$ (keep in mind that $\mu^2/M_P$ may be
small, but it is definitely finite). So to extract the correct behaviour of
(\ref{TrDet}) in the limit $Q \rightarrow 0$, one has to first expand in small $Q$ and
then expand this result in small $1/M_P$ to the desired order. Doing that, one
recovers (\ref{EigO}) at the zeroth order in the $Q$-expansion.

The lesson we learn from the above considerations is that (\ref{EigD}) is valid in a
neighborhood of the point $Q = \mu$, in which $Q$ and $\mu$ have comparable orders of magnitude. In that neighborhood both $m^2_{\pm}$ are positive
definite except at $Q = \mu$, where one has:\footnote{Strictly speaking, (\ref{MMQM}) is also valid in the small neighborhood in which $Q-\mu \rightarrow 0$, i.e. when $Q-\mu <\!\!< \mu^2/M_P$. The argument is analogous to the one below eq. (\ref{EigD}). \label{FR}}
\be \label{MMQM}
m^2_+ = 2 h^2 \mu^2 N_m^2 + \,{\cal O} \!\left( \frac{1}{M_P^2} \right) \quad , \qquad
m^2_-= - \,\frac{2 W_0^2 N_e}{M_P^4} + \frac{2 h^2 \mu^6 N_e^2 (N_m - 1)}{M_P^4} \,\, .
\ee
So, in principle, the sign of $m_-^2|_{Q=\mu}$ depends on the relative magnitudes of $W_0$ and $h \mu^3$, except for the case $N_m = 1$. However, if one wants the positive vacuum energy density in the metastable state to be very small, then the following relation should be satisfied \cite{AHKO, DPP}:
\be \label{W0hm}
\frac{W_0^2}{M_P^2} \approx h^2 \mu^4 N_e \, ,
\ee
where \,$\approx$ \,means that the two sides are of the same order of magnitude. If one
assumes this relation, then the first term in $m_-^2$ is dominant as by order of magnitude
it is $\sim h^2 \mu^4/M_P^2$ and so $m_-^2|_{Q=\mu} <0$. Hence, when (\ref{W0hm}) holds,
the new effect due to $W_0\neq 0$ corrects the order $1/M_P^2$ results instead of those at
${\cal O}(1/M_P^4)$.

Let us see what conclusions one can make for the sign of $m_-^2$ throughout the interval
$[0, \mu]$ for the cases when the $W_0$ contribution is of order $1/M_P^4$ and of order
$1/M_P^2$, respectively. In either case, the sign of the eigenvalue $m_-^2$ is determined
by the sign of ${\rm det} M$, see (\ref{TrDet}).\footnote{If ${\rm det} M < 0$, then the
expression under the square-root is greater than $({\rm Tr} M)^2$ and so $m_-^2 < 0$.}
However, in the first case, the leading order contribution is $\sim Q^2 (Q^2 - \mu^2)^2$
and so is positive-definite except at $Q=0$ and at $Q =\mu$.\footnote{More precisely, except
in very small neighborhoods of those points; see the remark in footnote \ref{FR}.} At
these two points the subleading term (of order $1/M_P^4$) determines the sign and it is
negative due to the negative $W_0$ contribution. On the other hand, in the case when
(\ref{W0hm}) holds we should only look at the terms of up to ${\cal O}(1/M_P^2)$. The
leading contribution to ${\rm det} M$, which is first order in $1/M_P^2$, is now
${\rm det} M^{(1)} \sim Q^2 (Q^2 - \mu^2)^2 h^2 N_m N_e - Q^2 W_0^2 N_e /M_P^2$, which
upon using (\ref{W0hm}) gives
\be
{\rm det} M^{(1)} \approx Q^2 h^2 [(Q^2 - \mu^2)^2 N_m N_e - \mu^4 N_e^2] < 0
\ee
at any point in $(0, \mu]$ as $N_m < N_e$. To recapitulate: when (\ref{W0hm}) is imposed,
we find that $m_-^2 < 0$ for any $Q \in [0, \mu]$, while without (\ref{W0hm}) $m_-^2$ is
negative only in small neighborhoods of $Q=0$ and $Q=\mu$. In a similar way one can show 
that $m_+^2$ is positive-definite without (\ref{W0hm}), whereas with (\ref{W0hm}) it is 
negative only in a small neighborhood of the origin.

The above conclusions imply that at some tempreature $T_c^{\varphi}$, below $T_c^Q$ of
Section 5, there will be another second order phase transition, this time in the meson
direction corresponding to $m_-^2$. We will explain in Section \ref{Disc} that this 
phase transition is actually necessary in order for the system to roll towards the 
metastable vacuum, due to the shifting of the latter from $(\varphi_1,\varphi_2) = (0,0)$ 
to $(\varphi_1,\varphi_2) \sim (\mu^2/M_P,\mu^2/M_P)$. However, computing the critical 
temperature is now much more complicated than in the quark case.
One reason is that the phase transition occurs at some $Q\neq 0$. (With (\ref{W0hm})
imposed, this could apriori be any point in the interval $[0,\mu]$, whereas without this
constraint it has to occur at $Q=\mu$.) Another, much more serious issue is that for
temperatures below $T_c^Q$ there are masses that are greater than the temperature
and so the high-temperature approximation (\ref{Texpansion}) cannot be used. Instead, one
should consider the full integral expression:
\be
V_{eff} (\hat{\chi}) = V_0 (\hat{\chi}) + \frac{T^4}{2\pi^2} \sum_I \pm n_I \int_0^{\infty}
dx \, x^2 \ln \!\left( 1 \mp \exp ( - \sqrt{x^2 + m^2_I (\hat{\chi})/T^2} ) \right),
\ee
where $n_I$ are the numbers of degrees of freedom and the upper (lower) sign is for
bosons (fermions).

Unfortunately, that means that we cannot obtain a simple analytic
answer for the meson critical temperature $T_c^{\varphi}$.
However, we can make an estimate of its magnitude in the vein of
\cite{ACJK}. Namely, let us consider a path in field space
connecting the point along the Q-axis, at which the rolling in the
meson direction starts, with the point
$(Q,\varphi_1,\varphi_2)\neq (0,0,0)$, where the metastable
minimum is, and take into account only fields whose masses change
significantly along this path.\footnote{Recall that the reason for
this is that fields with (nearly) constant masses contribute only
to the field-independent $T^4$ term in the effective potential,
which we are not interested in.} Now, let us assume that all
relevant masses are either much smaller or much greater than $T$.
We will see below that our results are consistent with this
assumption. Then for fields with $m<\!\!<T$ we can still use
(\ref{Texpansion}), whereas for fields with $m>\!\!>T$ we can
utilize the approximation (see (3.5) of \cite{ACJK}): 
\be \label{Approx} 
\pm \int_0^{\infty} dx \, x^2 \ln \!\left( 1 \mp
\exp (- \sqrt{x^2+m^2/T^2}) \right) \sim T^4 \left( \frac{m}{2\pi
T} \right)^{3/2} \exp \left( - \frac{m}{T} \right). 
\ee 
In principle, the full temperature-dependent part of the effective
potential is obtained by summing over all fields. Note however,
that the exponential in (\ref{Approx}) strongly suppresses the
contributions of the fields with mass $m>\!\!>T$. In other words,
as long as there is at least one field with $m<\!\!<T$, one can
neglect the heavy fields to leading order.

As we saw above, without imposing (\ref{W0hm}) the phase
transition occurs at $Q=\mu$. Let us assume that this is true also
with (\ref{W0hm}). Clearly, for the latter case this assumption
will only give us a lower bound on $T_c^{\varphi}$, but this is as
good as one can get in that case without studying the full
dynamical evolution of the system. At $Q=\mu$, the heavy masses in
our system are $m^2_{Re q_{1+}}$, $m^2_{Im q_{1+}}$, $m^2_{Re
q_{2-}}$, $m^2_{Im q_{2+}}$ and $m^2_+$. All of them are \,$\sim
\,h^2 \mu^2 + \,{\cal O}(1/M_P^2)$ and to leading order remain
constant along the above field space path. The light masses that
determine $T_c^{\varphi}$ in this approximation are $m^2_{Re
q_{1-}}$, $m^2_{Im q_{1-}}$, $m^2_{Re q_{2+}}$, $m^2_{Im q_{2-}}$
and $m^2_-$ .{}\footnote{All of them vary significantly (meaning
that the magnitude of their change is comparable to the magnitude
of their leading order) along the path of interest and so
contribute essentially to the effective potential.} Hence we can
obtain an estimate for the critical temperature by solving 
\be
\pd^2_{\varphi_-} V_T^{l} = - m^2_- \quad , \quad \pd_{\varphi_-}
V_T^{l} = - \pd_{\varphi_-} V_0 \qquad {\rm at} \qquad
(Q,\varphi_1,\varphi_2)=(\mu,\varphi_1^*,\varphi_2^*) \, . 
\ee 
In this system of equations $V_T^{l}$ denotes the
temperature-dependent part of the effective potential, that
results from (\ref{Texpansion}) by summing only over the light
fields, and $\varphi_-$ is the linear combination of $\varphi_1$
and $\varphi_2$ that corresponds to the mass-squared eigenvalue
$m^2_-$. One then finds that 
\be \label{CritTp}
T_c^{\varphi} \sim \frac{W_0}{h M_P^2} \approx \frac{\mu^2}{M_P} \, , 
\ee 
where in the second equality we have used (\ref{W0hm}). Note that the 
heavy masses (let us denote them by $m_h$ with $h$ running over all 
heavy fields) are all of zeroth order in $1/M_P$, whereas the light 
masses (denote them by $m_l$ with $l$ running over the light fields) 
are all multiplied by the small constant $h$ compared to (\ref{CritTp}). 
So our assumption, that at temperatures near $T_c^{\varphi}$ we have 
$m_h>\!\!>T>\!\!>m_l$, is consistent with the estimate in (\ref{CritTp}).

Before concluding this section, let us comment on the first order
phase transition found in \cite{FKKMT}. Its critical temperature
is of order $(\hat{T}_c^{\varphi})^2 \sim h \mu^2$. When it is reached,
tunnelling becomes possible between $\la \Phi \ra = 0$ and a minimum away
from the origin in the meson direction $\la \Phi \ra \sim \unit_{N_f}$,
which at zero temperature becomes the supersymmetric vacuum at
$\langle\Phi\rangle=h^{-1}\mu \, \epsilon^{(3N-N_f)/{(N_f-N)}}
\,\unit_{N_f}$. On the other hand,
our second order phase transition occurs in the field direction
$\varphi_- = \varphi_2 + \frac{N_e}{N_m} \frac{\mu^2}{M_P^2} \varphi_1 + 
{\cal O}(\frac{1}{M_P^4})$ and at $\la q_1 \ra = \la \tilde{q}_1 \ra\neq 
0$.\footnote{This expression for $\varphi_-$ is valid with or without 
(\ref{W0hm}) as $W_0$ appears for first time at order $1/M_P^6$.}
Which one of $T_c^{\varphi}$ and $\hat{T}_c^{\varphi}$
is greater depends on the relative magnitudes of $h$ and
$(\mu/M_P)^2$. However, regardless of that, the second order phase
transition is much more likely to take precedence since the first
order one, as any tunnelling event, is exponentially suppressed.
To gain a better understanding of the phase structure in
the meson direction and, in particular, to be able to estimate the
supergravity corrections to the height of the potential barrier relevant
for the first order phase transition, we would need to take into
account the non-perturbative dynamically generated contribution to
the superpotential. We leave that for future research.

\section{Towards KKLT-ISS at finite $T$} \label{KKLTISS}
\setcounter{equation}{0}

The proposal of \cite{KKLT} is a significant progress towards
finding dS vacua in string theory with all moduli stabilized.
However, the uplifting of their AdS minimum to a de Sitter one has
been rather difficult to implement in a controlled way. It was
shown recently in \cite{DPP, AHKO}, that this can be achieved
easily by using the ISS model as the uplifting sector.
They considered the following coupling:
\be \label{WK}
W = W_{ISS} + W_{KKLT} \, , \qquad K = K_{ISS} + K_{KKLT} \, ,
\ee
where $W_{ISS}$ and $K_{ISS}$ are given by (\ref{Wtree}) and
(\ref{KaP}), whereas:
\be
W_{KKLT} = \tilde{W}_0 + a e^{-b \rho} \qquad {\rm and} \qquad
K_{KKLT} = -3\ln (\rho + \bar{\rho}) \, .
\ee
In the string context, the constant $\tilde{W}_0$ is due to
nonzero background fluxes. By tuning it suitably, one can achieve
an almost vanishing cosmological constant and a light gravitino
mass \cite{DPP}, which is an important improvement compared to
the models with D-term uplifting. Therefore, it would be quite
interesting to investigate the phase structure of the KKLT-ISS
model (\ref{WK}) at finite temperature. In this section, we limit
ourselves to a discussion of the fate of the $(Q, \varphi_1,
\varphi_2) = (0,0,0)$ ISS minimum when the KKLT field $\rho$ is
included.

In Subsection \ref{local}, we noted that the presence of terms
linear in the meson vevs $\varphi_1$ and $\varphi_2$ can shift
the minimum of the effective potential away from the origin of the
ISS field space. This, of course, refers to terms that are not
multiplying $Q^2$ since $Q=0$ is always a local minimum as long as
$\mu <\!\!< M_P$, see eq. (\ref{VTQ2}). However, in
that section such linear terms were not appearing. Now the situation
is very different, as the inclusion of the KKLT sector introduces
many terms linear in $\varphi_1$ and $\varphi_2$. One source of them
comes from the constant piece $\tilde{W}_0$ in the KKLT
superpotential, as one can easily convince oneself by looking at
the form of $\la G^I G_I \ra$ in (\ref{GIGI}).\footnote{Recall that
the full expression is $\la e^G G^I G_I \ra$. Hence the mixed terms
in (\ref{GIGI}) give contributions proportional to $\varphi_1
\tilde{W}_0$ and to $\varphi_2 \tilde{W}_0$.} In
addition, there are many more contributions, linear in the
meson vevs, from mixed terms between the KKLT and ISS sectors in the
total Tr$M^2$. For example, the total fermionic Tr$M_f^2$ is now
\be
{\rm Tr} M_f^2 = \la e^G \left[ K^{A \bar{B}} K^{C \bar{D}} (\nabla_A G_C
+ G_A G_C) (\nabla_{\bar{B}} G_{\bar{D}} + G_{\bar{B}}
G_{\bar{D}}) - 2 \right] \ra \, ,
\ee
where $A = \{\rho, I\}$ and the index $I$ runs over the ISS fields.
Writing this out, we have
\bea \label{TrMKKLT}
{\rm Tr} M_f^2 &=& \la e^G \left[ K^{\rho \bar{\rho}} K^{\rho
\bar{\rho}} (\nabla_{\rho} G_{\rho} + G_{\rho} G_{\rho})
(\nabla_{\bar{\rho}} G_{\bar{\rho}} + G_{\bar{\rho}}
G_{\bar{\rho}}) \right. \nn \\
&+& 2 K^{\rho \bar{\rho}} K^{I \bar{J}} (\nabla_{\rho} G_{I} + G_{\rho} G_{I})
(\nabla_{\bar{\rho}} G_{\bar{J}} + G_{\bar{\rho}} G_{\bar{J}}) \nn \\
&+& \left. (G_{IJ} + G_I G_J) (G^{IJ} + G^I G^J) - 2 \right] \ra \, ,
\eea
where in the second line we have used that $\nabla_{\rho} G_I
= \nabla_I G_{\rho}$.\footnote{The relation $\nabla_{\rho} G_I
= \nabla_I G_{\rho}$ is due to $\pd_I G_{\rho} = \pd_{\rho} G_I =
- W_{\rho} W_I /W^2$ together with the fact that the only
nonvanishing Christoffel symbols are $\Gamma^{\rho}_{\rho \rho}$
and $\Gamma^{\bar{\rho}}_{\bar{\rho} \bar{\rho}}$.}
The last line is
the familiar ISS plus supergravity contribution, which for $W =
W_{ISS} +W_{KKLT}$ gives the linear term mentioned in the beginning
of this paragraph. Let us
now take a more careful look at the remaining terms. One can
verify that the second line contains, among many others, the terms
\be
\la \,\frac{3}{(\rho + \bar{\rho})^3} [W K_I \overline{W}^I +
\overline{W} K^I W_I ] \,\ra \, .
\ee
It is easy to see that these also
lead to contributions which are linear in $\varphi_1$ and
$\varphi_2$. Further linear terms come from the first line of
(\ref{TrMKKLT}) and also from the mixed terms in the scalar potential
\be
V = e^K (K^{A \bar{B}} D_A W D_{\bar{B}} \overline{W} - 3 |W|^2)
\ee
which determines the bosonic Tr$M_b^2$ via \cite{BG2}:
\be
{\rm Tr} M^2_b = 2 \,\la K^{C\bar{D}} \,\frac{\partial^2 V}{\partial \chi^C
\partial \bar{\chi}^{\bar{D}}} \,\ra \, .
\ee

Therefore, in the coupled KKLT-ISS model,
generically, the origin of the ISS field space $(Q, \varphi_1,
\varphi_2) = (0,0,0)$ is not a local minimum anymore. The
coefficients of the terms linear in the meson vevs are functions
of the KKLT field $\rho$. Hence $\la \rho \ra$ is related to the
magnitude of the shift of the minimum in the ISS plane. Although
very interesting, we leave further analysis of the KKLT-ISS
system for future work.

\section{Discussion} \label{Disc}
\setcounter{equation}{0}

We studied the effective potential at finite temperature for the ISS model
coupled to supergravity. Assuming that at high temperature the fields are
at the origin of field space, which is a local minimum, we
investigated the phase structure of the system as it cools down. In the quark 
direction, the situation is analogous to the rigid case \cite{FKKMT}.
Namely, there is a second order phase transition at certain critical
temperature, $T_c^Q$. The effect of the supergravity corrections is to
decrease $T_c^Q$ compared to its global supersymmetry counterpart. In the
meson branch however, a new feature appears whenever the superpotential contains 
a nonvanishing constant piece $W_0$. Recall that in the global theory all meson 
fields always had a local minimum at the origin of the meson direction, with no
tree-level contributions to their masses-squared. Now we find that, when 
$W_0\neq 0$, for some of them this ceases to be true at some temperature
$T_c^{\varphi}$, below which negative tree-level supergravity corrections
to their effective masses-squared are outweighing the positive one-loop
temperature dependent contributions. Hence, the supergravity interactions lead
to the occurrence of a new second order phase transition whenever $W_0\neq 0$.

Since $T_c^{\varphi} < T_c^Q$, as we saw in Section \ref{mesonPhtr}, the fields
first start rolling away from the origin in the quark direction. When the temperature
decreases enough, the same happens also in the meson direction. However, unlike in 
the rigid case, the second phase transition does not imply that the system is moving 
away from the supersymmetry breaking vacuum. The reason is that the coupling to 
supergravity leads to slight shifting of the position of the metastable
minimum \cite{AHKO}.\footnote{This is not a trivial consequence of including 
supergravity, as in the ISS model coupled to sugra with $W_0 = 0$ there is 
{\it no} shifting of the metastable vacuum compared to the global case.}
Whereas in the rigid theory it was given by $\la q_1 \ra
= \la \tilde{q}_1 \ra = \mu \,\unit_{N_m}$ and $\la \Phi \ra = 0$, in the locally
supersymmetric case some of the meson vevs also acquire nonzero value:
$(\la diag \,\phi_{11} \ra , \la diag \,\phi_{22} \ra) = (\varphi_1, \varphi_2) \sim 
(\mu^2 / M_P , \mu^2 / M_P) <\!\!< \la q_1 \ra$. Hence the latter phase transition is,
in fact, a necessary condition for the system to evolve towards the metastable vacuum. 
Of course, to follow with more precision the evolution of this system as the temperature 
decreases, one has to study the full effective potential for both $\la q \ra \neq 0$ 
and $\la \Phi \ra \neq 0$ away from a small neighborhood of the origin.

In the above paragraph, we reached the conclusion that the final state of
the system at $T\sim 0$ is likely to be the metastable vacuum. However, one
should be cautious since, similarly to \cite{FKKMT}, our considerations assume thermal 
equilibrium. Hence, although suggestive, they are not completely
conclusive. As was pointed out in \cite{FKKMT}, for proper understanding of
the evolution of the system one should address also the dynamics of the fields
at finite temperature. Actually, even before worrying about dynamics, one may be
concerned that in our case the situation is complicated by the existence
of new supergravity-induced supersymmetric minima.\footnote{That is, susy minima
other than those induced by the non-perturbative superpotential $W_{ADS}$, see
eq. (\ref{WADS}). Recall that the Witten index gives only the number of global
susy minima of globally supersymmetric theories. Hence, in the present context it
is not applicable and so one cannot immediately rule out the presence of additional
solutions.} Indeed, it was shown in
\cite{DPP} that such a minimum occurs in the KKLT-ISS model.\footnote{It is
true that this minimum is much further out in field space than the metastable
one, but its very existence raises the possibility that it may be quite
premature to make conclusions about the final state at low temperature,
based solely on studying the immediate neighborhood of the origin.} However,
this new vacuum only appears due to the interaction with the KKLT sector; it
is easy to see that the last condition in the solution for this minimum, eq.
(19) of \cite{DPP}, is only satisfied with an appropriate choice of (some of)
the tunable KKLT parameters $W_0$, $a$ and $b$. Still, one may wonder whether
there could be a solution to the supersymmetry preserving equations for the ISS
plus supergravity sector alone. In Appendix B we show that this is not possible
(at least in the field-space drections of interest, i.e for an ansatz for the vevs
that is of the same type as the one in \cite{DPP}).

The new solution of \cite{DPP} is only one indication that the KKLT-ISS system
is quite intricate to study. Another is the fact that, as we saw in Section
\ref{KKLTISS}, the interaction with the KKLT sector leads to shifting of the
high-temperature minimum of the effective potential away from the origin of the
ISS field space. Understanding the phase structure of this system is of
great interest. However, the technical complications involved are rather
significant. Therefore, it may be beneficial to gain preliminary intuition
about it by considering the recently proposed O'KKLT model \cite{KL}, as the
latter is much simpler while having quite similar behavior. In addition, the
O'KKLT model has a significance of its own, as it was argued in \cite{KL} to
be of value in studies of cosmological inflation.\footnote{In fact, this
statement only applies to a variant of O'KKLT considered in Section 3 there.
This same variant is also the model that is most useful as an approximation
of KKLT-ISS since it is the one, whose supersymmetric global minimum is at
finite field values.} We hope to address this in the future.

\section*{Acknowledgements}
We would like to thank V. Khoze, R. Russo and G. Travaglini for useful
discussions. The work of L.A. is supported by the EC Marie Curie Research
Training Network MRTN-CT-2004-512194 {\it Superstrings}.

\appendix

\section{Useful mass matrix formulae}
\setcounter{equation}{0}

By differentiating the F-term part of (\ref{potMp}) w.r.t. to $\chi^I$ and
$\bar{\chi}_J$ one finds:
\be \label{Ap1}
\pd_I \pd^J V = e^G [M_P^6(G^{LJ}+ G^L
G^J)(G_{LI} + G_L G_I) - M_P^4 G_I G^J + \delta_I^J ( M_P^4 G^L G_L -2 M_P^2)] \, .
\ee
Substituting $e^G = e^{K/M_P^2} |W|^2/M_P^6$ and
\be
G_{LI} + G_L G_I = \frac{1}{M_P^2} \frac{W_{LI} + K_L W_I + K_I W_L}{W} +
\frac{K_L K_I}{M_P^4} \, ,
\ee
we see that all powers of $W$ in the denominator cancel out. We want to compute
the value of the resulting expression for two cases: One is for zero background vevs of
all scalars. And the other is for $\varphi_1, \varphi_2 = 0$ but $Q\neq 0$.

In the first case, i.e. for $Q, \varphi_1, \varphi_2 = 0$, one has that $\la K \ra = 0$
and $\la K_I \ra = 0$ for $\forall$ \!\!$I$. Let us denote $\la W \ra \equiv {W}_0$ and
for the moment consider $W_0 \neq 0$ for more generality. We find that
\be
\la \pd_I \pd^J V \ra = \la W_{IL} \overline{W}^{JL} \ra + \frac{1}{M_P^2} \left( \delta_I^J \la W_L \overline{W}^L \ra -
\la W_I \overline{W}^J \ra \right) \, ,
\ee
where all higher orders in the $1/M_P$ expansion vanish due to $\la K \ra,
\la K_I \ra = 0$ regardless of the value of $W_0$.
Realizing that the vevs of all double derivatives of $W$ vanish at the origin
of field space, we finally obtain:
\be
\label{massqbq}
\la \pd_I \pd^J V \ra = \frac{1}{M_P^2} \left( \delta_I^J \la W_L \overline{W}^L \ra - \la W_I
\overline{W}^J \ra \right) \, .
\ee
Note that this result, apart from being exact to all orders, is also completely
independent of $W_0$.

Applying (\ref{massqbq}) to compute the mass matrices
$m^2_{q \bar{q}}$\,, where $q$ is either $q_1$ or $\tilde{q}_1$
and $\bar{q}$ is either $\bar{q}_1$ or $\bar{\tilde{q}}_1$, one
arrives at half of the relations in (\ref{mixedMM}). The other half,
i.e. of types $m^2_{qq}$ and $m^2_{\bar{q} \bar{q}}$, can be derived
in a similar way. (It is perhaps more convenient to differentiate the
potential in the form (\ref{pot}).) One finds that at the origin of field space:
\be \label{massqq}
\la \pd_K \pd_L V \ra = \la K^{I \bar{J}} W_{KLI} \overline{W}_{\bar{J}}
\ra \, ,
\ee
which is exactly the expression for the rigid case, since
all second derivatives of the superpotential vanish for zero
background fields. And again, this is exact result to all orders in $1/M_P$.

In the case of $\varphi_1, \varphi_2 = 0$ but $Q\neq 0$, we will be interested in the
meson mass matrices. For $I,J$ running over the meson field components only, we still
have that $\la K_I \ra = 0$, although now $\la K \ra \neq 0$. Keeping $W_0\neq 0$ and
expanding (\ref{Ap1}) up to order $1/M_P^4$, we find:
\bea \label{ApMM}
\!\la \pd_I \pd^J V \ra \!\!\!&=& \!\!\!\la W_{IL} \overline{W}^{JL} \ra \\
\!\!\!&+& \!\!\!\frac{1}{M_P^2} \left\{
\delta_I^J \la W_L \overline{W}^L \ra - \la \overline{W}_I W^J \ra
+ \la W_I K_L \overline{W}^{JL} + W_{IL} K^L \overline{W}^J +
K W_{IL} \overline{W}^{JL} \ra \right\} \nn \\
\!\!\!&+& \!\!\!\frac{1}{M_P^4} \left\{ \!\delta_I^J \!\!\left( \!\!\la K W_L
\overline{W}^L \!\ra -
\!2 W_0^2 \!\right) \!+ \la K \ra \la W_I K_L \overline{W}^{JL} \!\!+ \!W_{IL} K^L
\overline{W}^J \!\ra + \la \frac{K^2}{2} W_{IL} \overline{W}^{JL} \!\ra \!\right\} \!\!. \nn
\eea
Obviously, for $Q=0$ the above formula agrees with (\ref{massqbq}). However, recall
that the $I,J$ indices in it run only over the meson fields, whereas $L$ runs also
over the squarks.

Finally, one can easily convince oneself that if $K,L$ run only over
the mesons, whereas the squarks are the only fields with nonzero vevs, then
$\la \pd_K \pd_L V \ra = 0$ to all orders and for any value of $W_0$.

\section{On existence of new susy solutions}
\setcounter{equation}{0}

We investigate here whether there are solutions to the supersymmetry preserving
equations for the ISS model coupled to supergravity, with the non-perturbative
superpotential $W_{ADS}$ of eq. (\ref{WADS}) still neglected.

The susy equations are:
\be \label{susyEq}
D_{q_1} W = 0 \, , \qquad D_{\tilde{q}_1} W = 0 \, , \qquad D_{\Phi_{11}} W = 0
\, , \qquad D_{\Phi_{22}} W = 0 \, .
\ee
Similarly to \cite{DPP}, we consider the following ansatz for the expectation
values of the quark and meson fields in the solutions we are seeking:
\be
\la q_1 \ra = \mu_1 \unit_{N_m} \, , \qquad \la \tilde{q}_1 \ra = \mu_2
\unit_{N_m} \, , \qquad \la \Phi_{11} \ra = \nu_1 \unit_{N_m} \, , \qquad
\la \Phi_{22} \ra = \nu_2 \unit_{N_e} \, ,
\ee
where all vevs are real. Evaluating (\ref{susyEq}) in this
background gives:
\bea \label{susyEq1}
D_{q_1} : \,\,\,\, h \mu_2 \nu_1 + \mu_1 \la W \ra = 0 \, &,& \qquad
D_{\tilde{q}_1} : \,\,\,\, h \mu_1 \nu_1 + \mu_2 \la W \ra = 0 \, , \nn \\
D_{\Phi_{11}} : \,\,\,\, h (\mu_1 \mu_2 - \mu^2) + \nu_1 \la W \ra = 0 \, &,&
\qquad \!\!D_{\Phi_{22}} : \,\,\,\, -h \mu^2 + \nu_2 \la W \ra = 0 \, .
\eea
The two equations on the first line of (\ref{susyEq1}) imply that
$\mu_1^2 = \mu_2^2$. On the other hand, the ones on the second line
lead to:
\be \label{nu2}
\nu_2 = - \frac{\mu^2}{\mu_1 \mu_2 - \mu^2} \, \nu_1 \, .
\ee
Now, using this last relation and the equations for $D_{q_1}$ and
$D_{\Phi_{22}}$, we find:\footnote{Note that, up to now, we have not used the
explicit form of $\la W \ra$. So our
results (\ref{nu}), together with $\mu_1^2 = \mu_2^2$\,, are valid also for the
KKLT-ISS set-up considered in \cite{DPP}. In fact, the solution given in their
eq. (19) is only valid for $\mu_1 = \mu_2$\,, in which case one can see that it
agrees with (\ref{nu}).}
\be \label{nu}
\nu_1^2 = \frac{\mu_1}{\mu_2} \left(\mu_1 \mu_2 - \mu^2\right) \,\, , \qquad
\nu_2 = - \mu^2 \left(\frac{\mu_1}{\mu_2 \left(\mu_1 \mu_2 - \mu^2\right)}
\right)^{1/2} \, .
\ee
So far, we have expressed all vevs in terms of one of them, which could be
either $\mu_1$ or $\mu_2$. Let us choose this to be $\mu_1$. To obtain an
independent equation for it, we need to use the explicit vev of the
superpotential:
\be
\la W \ra = h \left(\mu_1 \mu_2 \nu_1 N_m - \mu^2 (\nu_1 N_m + \nu_2 N_e)\right) \, .
\ee
Combining this with (\ref{nu2}) and the $D_{q_1}$ equation in (\ref{susyEq1}), we
find:
\be \label{Eqmu}
- \frac{\mu_2}{\mu_1} = \frac{1}{M_P^2} \left[ \mu_1 \mu_2 N_m - \mu^2
\left( N_m - \frac{\mu^2 N_e}{(\mu_1 \mu_2 - \mu^2)} \right) \right] ,
\ee
where we have reinserted the explicit dependence on $M_P$ that comes from
$D_I W = \pd_I W + (K_I/M_P^2) W$. Let us now consider first the case
$\mu_2 = \mu_1$. Then (\ref{Eqmu}) becomes a quadratic equation
for $\mu_1^2$, whose solutions are:
\be \label{mu_sol}
\left(\!\frac{\mu_1}{M_P}\!\right)^{\!\!2} = \left(\!\frac{\mu}{M_P}\!
\right)^{\!2} - \frac{1}{2 N_m} \pm \frac{1}{2 N_m} \sqrt{1 - 4 \left(
\!\frac{\mu}{M_P} \!\right)^{\!4} N_m N_e} \,\, .
\ee
Since we would like both $\mu_1 <\!\!< M_P$ and $\mu <\!\!< M_P$ in order to
have a reliable field theory description, only the plus sign in (\ref{mu_sol})
is meaningful. Hence, we have
\be
\left(\!\frac{\mu_1}{M_P}\!\right)^{\!\!2} = \left(\!\frac{\mu}{M_P}\!
\right)^{\!2} - \left( \!\frac{\mu}{M_P} \!\right)^{\!4} \!N_e + {\cal O}
\left( \frac{\mu^{\,8}}{M_P^{\,8}} \right).
\ee
This implies that $\mu_1^2 - \mu^2 < 0$, which is inconsistent with (\ref{nu})
since the vevs $\nu_1$ and $\nu_2$ are real.
If we take in turn $\mu_2 = - \mu_1$ and repeat the above steps, we end up with
\be
\left( \!\frac{\mu_1}{M_P} \!\right)^{\!2} = - \left( \!\frac{\mu}{M_P}
\!\right)^{\!2} + {\cal O} \left( \frac{\mu^4}{M_P^4} \right) ,
\ee
which is again inconsistent for real vevs. So we conclude that coupling to
supergravity does not increase the number of vacua of the ISS model.

\end{document}